# Checking Interval Properties of Computations

Alberto Molinari · Angelo Montanari · Aniello Murano · Giuseppe Perelli · Adriano Peron

**Abstract** Model checking is a powerful method widely explored in formal verification. Given a model of a system, e.g., a Kripke structure, and a formula specifying its expected behaviour, one can verify whether the system meets the behaviour by checking the formula against the model.

Classically, system behaviour is expressed by a formula of a temporal logic, such as LTL and the like. These logics are "point-wise" interpreted, as they describe how the system evolves state-by-state. However, there are relevant properties, such as those constraining the temporal relations between pairs of temporally extended events or involving temporal aggregations, which are inherently "interval-based", and thus asking for an interval temporal logic.

In this paper, we give a formalization of the model checking problem in an interval logic setting. First, we provide an interpretation of formulas of Halpern and Shoham's interval temporal logic HS over finite Kripke structures, which allows one to check interval properties of computations. Then, we prove that the model checking problem

Alberto Molinari and Angelo Montanari
Department of Mathematics and Computer Science,
University of Udine, Italy
E-mail: molinari.alberto@gmail.com; angelo.montanari@uniud.it

Aniello Murano
Department of Electronic Engineering and Information Technology,
University of Napoli, Italy
E-mail: aniello.murano@unina.it

Giuseppe Perelli
Department of Computer Science
Oxford University, UK
E-mail: perelli.gi@gmail.com

Adriano Peron
Department of Electronic Engineering and Information Technology,
University of Napoli, Italy
E-mail: adrperon@unina.it

for HS against finite Kripke structures is decidable by a suitable small model theorem, and we provide a lower bound to its computational complexity.

**Keywords** Interval Temporal Logic · Model Checking · Decidability · Computational Complexity

# 1 Introduction

A classical problem in hardware and software system design is to come up with automatic techniques to ensure reliability. In this context, formal methods have provided structures and algorithms that have been successfully applied in several domains. One of the most notable techniques is model checking, where a formal specification of the desired properties of the system is checked against a model of its behaviour [6, 7, 24, 28]. The solution of the model checking problem, and thus its precise complexity, relies on the particular computational model and specification language we consider. In finite-state system verification, systems are usually modelled as labelled state-transition graphs, or Kripke structures, while specifications are formulas of a suitable (point-based) linear or branching temporal logic. The first attempt in this direction goes back to the late '70s, when the use of the linear temporal logic LTL in program verification was proposed by Pnueli [22]. LTL allows one to reason about changes in the truth value of formulas in a Kripke structure over a linearly-ordered temporal domain, where each moment in time has a unique possible future. More precisely, one has to consider all possible paths in a Kripke structure and to analyse, for each of them, how proposition letters, labelling the states, change from one state to the next one along the path. The model checking problem for LTL turns out to be PSPACE-complete [7, 23, 27].

Propositional interval temporal logics provide an alternative setting for reasoning about time. They have been applied in a variety of computer science fields, including artificial intelligence (reasoning about action and change, qualitative reasoning, planning, and natural language processing), theoretical computer science (specification and verification of programs), and databases (temporal and spatio-temporal databases) [10]. Interval-based temporal logics take intervals as their primitive temporal entities. Such a choice gives them the ability to express temporal properties, such as actions with duration, accomplishments, and temporal aggregations, which cannot be dealt with in standard (point-based) temporal logics.

A prominent position among interval temporal logics is occupied by Halpern and Shoham's modal logic of time intervals (HS, for short) [12]. HS features one modality for each of the 13 possible ordering relations between pairs of intervals (the so-called Allen's relations [1]), apart from the equality relation. As an example, the condition: "the current interval meets an interval over which $p$ holds" can be expressed in HS by the formula $\langle A \rangle p$, where $\langle A \rangle$ is the (existential) HS modality for Allen relation *meet*. In [12], it has been shown that the satisfiability problem for HS interpreted over all relevant (classes of) linear orders is highly undecidable. Since then, a lot of work has been done on the satisfiability problem for HS fragments, which showed that undecidability rules over them [2, 14, 17]. However, meaningful exceptions exist, including the interval logic of temporal neighbourhood and the logic of sub-intervals [3–5, 19].

Here, we focus our attention on the model checking problem for HS. While the satisfiability problem for HS and its fragments has been extensively and systematically investigated in the literature [8], a little work has been done on model checking. The idea is to evaluate HS formulas on finite Kripke structures making it possible to check the correctness of the behaviour of the system with respect to meaningful interval properties. To this end, we interpret each finite path of a Kripke structure as an interval, and we define its labelling on the basis of the labelling of the states that compose it, according to the homogeneity assumption [25]. Formally, we will show that finite Kripke structures can be suitably mapped into interval-based structures, called abstract interval models, over which HS formulas can be interpreted. Since finite Kripke structures may have loops, (abstract) interval models have, in general, an infinite domain. In order to devise a model checking procedure for HS over finite Kripke structures, we prove a small model theorem showing that, given an HS formula $\psi$ and a finite Kripke structure $\mathcal{K}$, there exists a finite interval model which is equivalent to the one induced by $\mathcal{K}$ with respect to the satisfiability of $\psi$. The main technical ingredients are (i) the definition of a suitable equivalence relation over finite paths (sequences) in $\mathcal{K}$, which is parametric in the nesting depth of Allen's modalities $\langle B \rangle$ and $\langle E \rangle$ in $\psi$, and (ii) the proof that the resulting quotient structure is finite and equivalent to the one induced by $\mathcal{K}$ with respect to the satisfiability of $\psi$.

The rest of the paper is organised as follows. In Section 2, we introduce syntax and semantics of HS (over interval models), and we establish a suitable connection between finite Kripke structures and abstract interval models. In Section 3, we introduce the fundamental notion of $BE_k$-descriptor. Next, in Section 4, we prove the small model theorem. Then, in Section 5, we show that the model checking problem for HS over finite Kripke structures is EXPSPACE-hard. Finally, in Section 6, we briefly discuss related work. Conclusions and future work directions are given in Section 7.

## 2 Interval temporal logic and Kripke structures

In this section, we give syntax and semantics of Halpern and Shoham's interval temporal logic HS with respect to (abstract) interval models. Moreover, we provide a suitable mapping from Kripke structures to interval models that allows us to interpret HS formulas over Kripke structures and then to define the notion of interval-based model checking.

### 2.1 The interval temporal logic HS

An interval algebra to reason about intervals and their relative order was first proposed by Allen in [1]; then, a systematic logical study of interval representation and reasoning was done by Halpern and Shoham, who introduced the interval temporal logic HS featuring one modality for each Allen interval relation [12].

Table 1 depicts 6 of the 13 possible binary ordering relations between a pair of intervals. The other 7 are the equality and the 6 inverse relations (given a generic binary relation $\mathcal{R}$, the inverse relation $\overline{\mathcal{R}}$ holds between two elements, $b\overline{\mathcal{R}}a$, if and only if $a\mathcal{R}b$).

**Table 1** Allen's interval relations and corresponding HS modalities.

| Allen relation | HS modality | Definition w.r.t. interval structures | Example |
|---|---|---|---|
| MEETS | $\langle A \rangle$ | $[x,y]\mathcal{R}_A[v,z] \iff y = v$ | |
| BEFORE | $\langle L \rangle$ | $[x,y]\mathcal{R}_L[v,z] \iff y < v$ | |
| STARTED-BY | $\langle B \rangle$ | $[x,y]\mathcal{R}_B[v,z] \iff x = v \wedge z < y$ | |
| FINISHED-BY | $\langle E \rangle$ | $[x,y]\mathcal{R}_E[v,z] \iff y = z \wedge x < v$ | |
| CONTAINS | $\langle D \rangle$ | $[x,y]\mathcal{R}_D[v,z] \iff x < v \wedge z < y$ | |
| OVERLAPS | $\langle O \rangle$ | $[x,y]\mathcal{R}_O[v,z] \iff x < v < y < z$ | |

In the table, each Allen relation is shown together with the corresponding HS (existential) modality. In its original formulation, HS allowed point intervals as well, that is, intervals consisting of a single point, but that way HS modalities are neither *mutually exclusive* nor *jointly exhaustive*, i.e., more than one relation, or even none, may hold between any two intervals. In the following, we will consider only strict intervals, consisting of two or more points (*strict semantics*).

The language of HS features a set of proposition letters $\mathcal{AP}$, the Boolean connectives $\neg$ and $\wedge$, the logical constants $\top$ and $\bot$ (respectively *true* and *false*), and a temporal modality for each of the (non trivial) Allen's relations, namely, $\langle A \rangle, \langle L \rangle, \langle B \rangle, \langle E \rangle, \langle D \rangle, \langle O \rangle, \langle \overline{A} \rangle, \langle \overline{L} \rangle, \langle \overline{B} \rangle, \langle \overline{E} \rangle, \langle \overline{D} \rangle$, and $\langle \overline{O} \rangle$.

Formally, HS formulas are defined by the following grammar:

$$\psi ::= p \mid \neg \psi \mid \psi \wedge \psi \mid \langle X \rangle \psi \mid \langle \overline{X} \rangle \psi, \quad \text{with } p \in \mathcal{AP}.$$

In the following, we will make use of the standard abbreviations of propositional logic, e.g., we will write $\psi \vee \varphi$ for $\neg \psi \wedge \neg \varphi$, $\psi \rightarrow \varphi$ for $\neg \psi \vee \varphi$, and $\psi \leftrightarrow \varphi$ for $(\psi \rightarrow \varphi) \wedge (\varphi \rightarrow \psi)$. Moreover, for all $X$, dual universal modalities $[X]\psi$ and $[\overline{X}]\psi$ are respectively defined as $\neg \langle X \rangle \neg \psi$ and $\neg \langle \overline{X} \rangle \neg \psi$, as usual.

Finally, it can be easily shown that, when the strict semantics is assumed, all HS modalities can be expressed in terms of modalities $\langle A \rangle, \langle B \rangle, \langle E \rangle$, and the transposed modalities $\langle \overline{A} \rangle, \langle \overline{B} \rangle, \langle \overline{E} \rangle$ as follows:

$$\langle L \rangle \psi \equiv \langle A \rangle \langle A \rangle \psi \qquad \langle \overline{L} \rangle \psi \equiv \langle \overline{A} \rangle \langle \overline{A} \rangle \psi$$
$$\langle D \rangle \psi \equiv \langle B \rangle \langle E \rangle \psi \equiv \langle E \rangle \langle B \rangle \psi \qquad \langle \overline{D} \rangle \psi \equiv \langle \overline{B} \rangle \langle \overline{E} \rangle \psi \equiv \langle \overline{E} \rangle \langle \overline{B} \rangle \psi$$
$$\langle O \rangle \psi \equiv \langle E \rangle \langle \overline{B} \rangle \psi \qquad \langle \overline{O} \rangle \psi \equiv \langle B \rangle \langle \overline{E} \rangle \psi$$

Given any subset of Allen's relations $\{X_1, \cdots, X_n\}$, we denote by $HS[X_1, \cdots, X_n]$ the fragment of HS that features modalities $X_1, \cdots, X_n$ only. As an example, we denote by $HS[A, \overline{A}, B, \overline{B}, E, \overline{E}]$ the HS fragment that features modalities $\langle A \rangle, \langle \overline{A} \rangle, \langle B \rangle, \langle \overline{B} \rangle, \langle E \rangle$, and $\langle \overline{E} \rangle$ only (observe that this fragment contains an equivalent formula for every HS formula).

HS can be viewed as a multi-modal logic with six primitive modalities, namely, $\langle A \rangle, \langle B \rangle, \langle E \rangle$, and their inverses. Accordingly, HS semantics can be defined over a multi-modal Kripke structure, here called abstract interval model, in which (strict) intervals are treated as atomic objects and Allen's relations as simple binary relations between pairs of intervals.

**Definition 1** An *abstract interval model* is a tuple $\mathcal{A} = (\mathcal{AP}, \mathbb{I}, A_{\mathbb{I}}, B_{\mathbb{I}}, E_{\mathbb{I}}, \sigma)$, where:

- $\mathcal{AP}$ is a finite set of proposition letters;
- $\mathbb{I}$ is a possibly infinite set of atomic objects (worlds);
- $A_{\mathbb{I}}, B_{\mathbb{I}}, E_{\mathbb{I}}$ are three binary relations over $\mathbb{I}$;
- $\sigma : \mathbb{I} \mapsto 2^{\mathcal{AP}}$ is a (total) labeling function, which assigns a set of proposition letters to each world.

Intuitively, in the interval setting, $\mathbb{I}$ is a set of intervals, $A_{\mathbb{I}}$, $B_{\mathbb{I}}$, and $E_{\mathbb{I}}$ are interpreted as Allen's interval relations $A$ (*meets*), $B$ (*started-by*), and $E$ (*finished-by*), respectively, and $\sigma$ assigns to each interval the set of proposition letters that hold over it.

Given an abstract interval model $\mathcal{A} = (\mathcal{AP}, \mathbb{I}, A_{\mathbb{I}}, B_{\mathbb{I}}, E_{\mathbb{I}}, \sigma)$ and an interval $I \in \mathbb{I}$, the truth of an HS formula over $I$ is defined by induction on the structural complexity of the formula as follows:

- $\mathcal{A}, I \models p$ iff $p \in \sigma(I)$, for any proposition letter $p \in \mathcal{AP}$;
- $\mathcal{A}, I \models \neg \psi$ iff it is not true that $\mathcal{A}, I \models \psi$;
- $\mathcal{A}, I \models \psi \vee \varphi$ iff $\mathcal{A}, I \models \psi$ or $\mathcal{A}, I \models \varphi$;
- $\mathcal{A}, I \models \langle X \rangle \psi$, for $X \in \{A, B, E\}$, iff there exists $J \in \mathbb{I}$ such that $I X_{\mathbb{I}} J$ and $\mathcal{A}, J \models \psi$;
- $\mathcal{A}, I \models \langle \overline{X} \rangle \psi$, for $\overline{X} \in \{\overline{A}, \overline{B}, \overline{E}\}$, iff there exists $J \in \mathbb{I}$ such that $J X_{\mathbb{I}} I$ and $\mathcal{A}, J \models \psi$.

*Satisfiability* and *validity* are defined in the usual way: an HS formula $\psi$ is satisfiable if there exists an interval model $\mathcal{A}$ and a world (interval) $I$ such that $\mathcal{A}, I \models \psi$. Moreover, $\psi$ is valid, denoted as $\models \psi$, if $\mathcal{A}, I \models \psi$ for all worlds (intervals) $I$ of any interval model $\mathcal{A}$.

## 2.2 Kripke structures and abstract interval models

Finite state systems are usually modelled as finite Kripke structures. In the following, we first recall the definition of finite Kripke structure and then we define a suitable mapping from this class of structures to abstract interval models that makes it possible to specify properties of systems by means of HS formulas.

**Definition 2** (Finite Kripke structure) A finite Kripke structure is a tuple $\mathcal{K} = (\mathcal{AP}, W, \delta, \mu, w_0)$, where $\mathcal{AP}$ is a set of proposition letters, $W$ is a finite set of states (*worlds*), $\delta \subseteq W \times W$ is a left-total relation between pairs of states (*accessibility relation*), $\mu : W \mapsto 2^{\mathcal{AP}}$ a total labelling function, and $w_0 \in W$ is the initial state.

For all $w \in W$, $\mu(w)$ captures the set of proposition letters that hold at that state; $\delta$ is the transition relation that constrains the evolution of the system over time. The relation $\delta$ is left-total because the paths of $\mathcal{K}$ are meant to represent system computations.

A simple Kripke structure, consisting of two states only, is reported in the following example. We will use it as a running example in the rest of the paper.

*Example 1* Figure 1 below depicts a two-state Kripke structure $\mathcal{K}_{Equiv}$ (the initial state is identified by a double circle). Despite its simplicity, it features an infinite number of different (finite) paths. Formally, $\mathcal{K}_{Equiv}$ is defined by the following quintuple: $(\{p,q\}, \{v_0, v_1\}, \{(v_0,v_0), (v_0,v_1), (v_1,v_0), (v_1,v_1)\}, \mu, v_0)$, where $\mu(v_0) = \{p\}$ and $\mu(v_1) = \{q\}$.

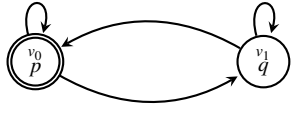

**Fig. 1** The Kripke structure $\mathcal{K}_{Equiv}$.

**Definition 3** (Track over $\mathcal{K}$) A track $\rho$ over a finite Kripke structure $\mathcal{K} = (\mathcal{AP}, W, \delta, \mu, w_0)$ is a finite sequence of states $v_0 \cdots v_n$, with $n \geq 1$, such that for all $i \in \{0, \cdots, n-1\}$, $(v_i, v_{i+1}) \in \delta$.

Let $\mathrm{Trk}_{\mathcal{K}}$ be the (infinite) set of all tracks over a finite Kripke structure $\mathcal{K}$. For any track $\rho = v_0 \cdots v_n \in \mathrm{Trk}_{\mathcal{K}}$, we define:

- $|\rho| = n+1$;
- $\rho(i) = v_i$;
- $\mathrm{states}(\rho) = \{v_0, \cdots, v_n\} \subseteq W$;
- $\mathrm{intstates}(\rho) = \{v_1, \cdots, v_{n-1}\} \subseteq W$;
- $\mathrm{fst}(\rho) = v_0$ and $\mathrm{lst}(\rho) = v_n$;
- $\rho(i,j) = v_i \cdots v_j$, $0 \leq i < j \leq |\rho|-1$ is a subtrack of $\rho$;
- $\mathrm{Pref}(\rho) = \{\rho(0,i) \mid 1 \leq i \leq |\rho|-2\}$ is the set of all proper prefixes of $\rho$;
- $\mathrm{Suff}(\rho) = \{\rho(i, |\rho|-1) \mid 1 \leq i \leq |\rho|-2\}$ is the set of all proper suffixes of $\rho$.

If $\mathrm{fst}(\rho) = w_0$, where $w_0$ is the initial state of $\mathcal{K}$, $\rho$ is said to be an *initial track*. Notice that the length of tracks, prefixes, and suffixes is greater than 1, as they will be mapped into strict intervals.

An abstract interval model (over $\mathrm{Trk}_{\mathcal{K}}$) can be naturally associated with a finite Kripke structure by interpreting every track as an interval bounded by its first and last states.

**Definition 4** (Abstract interval model induced by $\mathcal{K}$) The abstract interval model induced by a finite Kripke structure $\mathcal{K} = (\mathcal{AP}, W, \delta, \mu, w_0)$ is the abstract interval model $\mathcal{A}_{\mathcal{K}} = (\mathcal{AP}, \mathbb{I}, A_{\mathbb{I}}, B_{\mathbb{I}}, E_{\mathbb{I}}, \sigma)$, where:

- $\mathbb{I} = \mathrm{Trk}_{\mathcal{K}}$,
- $A_{\mathbb{I}} = \{(\rho, \rho') \in \mathbb{I} \times \mathbb{I} \mid \mathrm{lst}(\rho) = \mathrm{fst}(\rho')\}$,
- $B_{\mathbb{I}} = \{(\rho, \rho') \in \mathbb{I} \times \mathbb{I} \mid \rho' \in \mathrm{Pref}(\rho)\}$,
- $E_{\mathbb{I}} = \{(\rho, \rho') \in \mathbb{I} \times \mathbb{I} \mid \rho' \in \mathrm{Suff}(\rho)\}$,
- $\sigma : \mathbb{I} \mapsto 2^{\mathcal{AP}}$ such that for all $\rho \in \mathbb{I}$,

$$\sigma(\rho) = \bigcap_{w \in \mathrm{states}(\rho)} \mu(w).$$

In Definition 4, relations $A_\mathbb{I}, B_\mathbb{I}$, and $E_\mathbb{I}$ are interpreted as Allen's interval relations $A, B$, and $E$, respectively. Moreover, according to the definition of $\sigma$, a proposition letter $p \in \mathcal{AP}$ holds over $\rho = v_0 \cdots v_n$ if and only if it holds over all the states $v_0, \cdots, v_n$ of $\rho$. This conforms to the *homogeneity principle*, according to which a proposition letter holds over an interval if and only if it holds over all of its subintervals.

Satisfiability of an HS formula over a finite Kripke structure can be given in terms of induced abstract interval models.

**Definition 5** (Satisfiability of HS formulas over Kripke structures) Let $\mathcal{K}$ be a finite Kripke structure, $\rho$ be a track in $\text{Trk}_\mathcal{K}$, $\psi$ be an HS formula. We say that the pair $(\mathcal{K}, \rho)$ satisfies $\psi$, denoted by $\mathcal{K}, \rho \models \psi$, if and only if it holds that $\mathcal{A}_\mathcal{K}, \rho \models \psi$.

We are now ready to formally state the *model checking problem* for HS over finite Kripke structures: it is the problem of deciding whether $\mathcal{K} \models \psi$.

**Definition 6** Let $\mathcal{K}$ be a finite Kripke structure and $\psi$ be an HS formula. We say that $\mathcal{K}$ models $\psi$, denoted by $\mathcal{K} \models \psi$, if and only if

$$\text{for all \emph{initial} tracks } \rho \in \text{Trk}_\mathcal{K}, \text{ it holds that } \mathcal{K}, \rho \models \psi.$$

We conclude the section by giving some examples of meaningful properties of tracks that can be expressed in HS. To start with, we observe that the formula $[B]\bot$ can be used to select all and only the tracks of length 2. Indeed, given any $\rho$ with $|\rho| = 2$, independently of $\mathcal{K}$, it holds that $\mathcal{K}, \rho \models [B]\bot$, because $\rho$ has not (strict) prefixes. On the other hand, it holds that $\mathcal{K}, \rho \models \langle B \rangle \top$ if (and only if) $|\rho| > 2$. Modality $\langle B \rangle$ can actually be used to constrain the length of an interval to be greater than, less than, or equal to any value $k$. Let us denote $k$ nested applications of $\langle B \rangle$ by $\langle B \rangle^k$. It holds that $\mathcal{K}, \rho \models \langle B \rangle^k \top$ if and only if $|\rho| \geq k + 2$. Analogously, $\mathcal{K}, \rho \models [B]^k \bot$ if and only if $|\rho| \leq k + 1$. Let $\ell(k)$ be a shorthand for $[B]^{k-1}\bot \wedge \langle B \rangle^{k-2}\top$. It holds that $\mathcal{K}, \rho \models \ell(k)$ if and only if $|\rho| = k$.

Let us consider now the finite Kripke structure $\mathcal{K}_{Equiv}$ of Example 1, depicted in Figure 1. For the sake of brevity, for any track $\rho$, we denote by $\rho^n$ the track obtained by concatenating $n$ copies of $\rho$. The truth of the following statements can be easily checked:

- $\mathcal{K}_{Equiv}, (v_0v_1)^2 \models \langle A \rangle q$;
- $\mathcal{K}_{Equiv}, v_0v_1v_0 \not\models \langle A \rangle q$;
- $\mathcal{K}_{Equiv}, (v_0v_1)^2 \models \langle \overline{A} \rangle p$;
- $\mathcal{K}_{Equiv}, v_1v_0v_1 \not\models \langle \overline{A} \rangle p$.

The above statements show that modalities $\langle A \rangle$ and $\langle \overline{A} \rangle$ can be used to distinguish between tracks that start or end at different states.

Modalities $\langle B \rangle$ and $\langle E \rangle$ can be exploited to distinguish between tracks encompassing a different number of iterations of a given loop. This is the case, for instance, with the following statements:

- $\mathcal{K}_{Equiv}, (v_1v_0)^3 v_1 \models \langle B \rangle \big( \langle A \rangle p \wedge \langle B \rangle (\langle A \rangle p \wedge \langle B \rangle \langle A \rangle p) \big)$;
- $\mathcal{K}_{Equiv}, (v_1v_0)^2 v_1 \not\models \langle B \rangle \big( \langle A \rangle p \wedge \langle B \rangle (\langle A \rangle p \wedge \langle B \rangle \langle A \rangle p) \big)$.

Finally, HS makes it possible to distinguish between tracks $\rho_1 = v_0^3 v_1 v_0$ and $\rho_2 = v_0 v_1 v_0^3$, which involve the same number of iterations of the same loops, but differ in the order of loop occurrences: $\mathcal{K}_{Equiv}, \rho_1 \models \langle B \rangle (\langle A \rangle q \wedge \langle B \rangle \langle A \rangle p)$, but $\mathcal{K}_{Equiv}, \rho_2 \not\models \langle B \rangle (\langle A \rangle q \wedge \langle B \rangle \langle A \rangle p)$.

*Example 2* In Figure 2, we provide an example of a finite Kripke structure $\mathcal{K}_{Sched}$ that models the behaviour of a scheduler serving three processes which are continuously requesting the use of a common resource. The initial state is $v_0$: no process is served in that state. In any other state $v_i$ and $\bar{v}_i$, with $i \in \{1, 2, 3\}$, the $i$-th process is served (this is denoted by the fact that $p_i$ holds in those states). For the sake of readability, edges are marked either by $r_i$, for *request*$(i)$, or by $u_i$, for *unlock*$(i)$. Edge labels do not have a semantic value, that is, they are neither part of the structure definition, nor proposition letters; they are simply used to ease reference to edges. Process $i$ is served in state $v_i$, then, after "some time", a transition $u_i$ from $v_i$ to $\bar{v}_i$ is taken; subsequently, process $i$ cannot be served again immediately, as $v_i$ is not directly reachable from $\bar{v}_i$ (the scheduler cannot serve the same process twice in two successive rounds). A transition $r_j$, with $j \neq i$, from $\bar{v}_i$ to $v_j$ is then taken and process $j$ is served. This structure can easily be generalised to a higher number of processes.

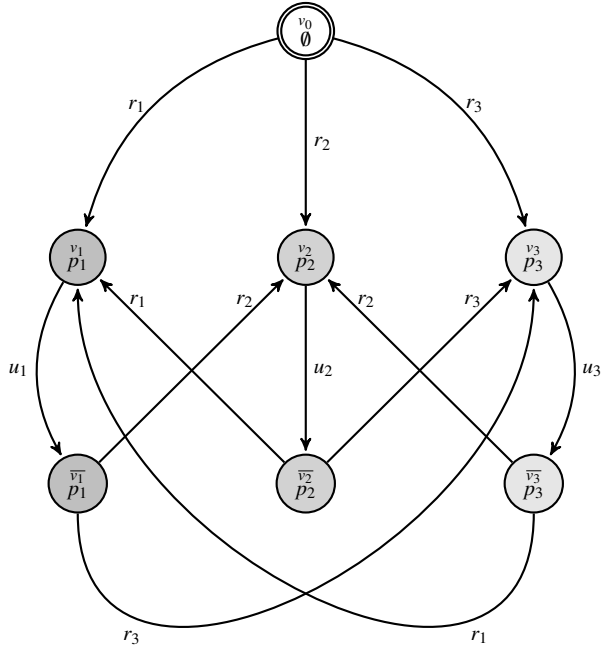

**Fig. 2** The Kripke structure $\mathcal{K}_{Sched}$.

We show how some meaningful properties to be checked over $\mathcal{K}_{Sched}$ can be expressed in HS. In all formulas, we force the validity of the considered property over all legal computation sub-intervals by using modality $[E]$ (all computation sub-intervals are suffixes of at least one initial track). Moreover, we will make use of the shorthand $wit_{\geq 2}(\{p_1,p_2,p_3\})$ for the formula:

$$(\langle D \rangle p_1 \wedge \langle D \rangle p_2) \vee (\langle D \rangle p_1 \wedge \langle D \rangle p_3) \vee (\langle D \rangle p_2 \wedge \langle D \rangle p_3),$$

which states that there exist at least two sub-intervals such that $p_i$ holds over the former and $p_j$ over the latter, with $i,j \in \{1,2,3\}$ and $j \neq i$ (such a formula can be easily generalised to an arbitrary set of proposition letters and to any natural number $k$). The truth of the following statements can be easily checked:

- $\mathcal{K}_{Sched} \models [E]\left(\langle B \rangle^5 \top \to wit_{\geq 2}(\{p_1,p_2,p_3\})\right)$;
- $\mathcal{K}_{Sched} \not\models [E]\left(\langle B \rangle^{10} \top \to \langle D \rangle p_3\right)$;
- $\mathcal{K}_{Sched} \not\models [E]\left(\langle B \rangle^7 \top \to \langle D \rangle p_1 \wedge \langle D \rangle p_2 \wedge \langle D \rangle p_3\right)$.

The first formula states that in any suffix of an initial track of length greater than or equal to 7 at least 2 proposition letters are witnessed. $\mathcal{K}_{Sched}$ satisfies the formula since a process cannot be executed twice in a row. The second formula states that in any suffix of an initial track of length at least 12 process 3 is executed at least once in some internal states. $\mathcal{K}_{Sched}$ does not satisfy the formula since the scheduler can avoid executing a process ad libitum. The third formula states that in any suffix of an initial track of length greater than or equal to 9, $p_1, p_2, p_3$ are all witnessed. The only way to satisfy this property is to constrain the scheduler to execute the three processes in a strictly periodic manner, but this is not the case.

## 3 The fundamental notion of $BE_k$-descriptor

In the previous section, we have shown that, for any given finite Kripke structure $\mathcal{K}$, one can find a corresponding induced abstract interval model $\mathcal{A}_\mathcal{K}$, featuring one interval for each track of $\mathcal{K}$. Since $\mathcal{K}$ has loops (each state must have at least one successor), the number of its tracks, and thus the number of intervals of $\mathcal{A}_\mathcal{K}$, is infinite. In this section, we prove that, given a finite Kripke structure $\mathcal{K}$ and an HS formula $\varphi$, there exists a *finite* abstract interval model, which is equivalent to $\mathcal{A}_\mathcal{K}$ with respect to the satisfiability of $\varphi$ (in fact, of a class of HS formulas including $\varphi$).

We start with the definition of some basic notions. The first one is the notion of BE-nesting depth of an HS formula.

**Definition 7** (BE-nesting depth of an HS formula) Let $\psi$ be an HS formula. The BE-nesting depth of $\psi$, denoted by $\text{Nest}_{BE}(\psi)$, is defined by induction on the structure complexity of the formula as follows:

- $\text{Nest}_{BE}(p) = 0$, for any proposition letter $p \in \mathcal{AP}$;
- $\text{Nest}_{BE}(\neg \psi) = \text{Nest}_{BE}(\psi)$;
- $\text{Nest}_{BE}(\psi \wedge \varphi) = \max\{\text{Nest}_{BE}(\psi), \text{Nest}_{BE}(\varphi)\}$;
- $\text{Nest}_{BE}(\langle B \rangle \psi) = \text{Nest}_{BE}(\langle E \rangle \psi) = 1 + \text{Nest}_{BE}(\psi)$;
- $\text{Nest}_{BE}(\langle X \rangle \psi) = \text{Nest}_{BE}(\psi)$, for $X \in \{A, \overline{A}, \overline{B}, \overline{E}\}$.

Making use of the notion of BE-nesting depth of a formula, we can define a relation of *k*-equivalence over tracks.

**Definition 8** Let $\mathcal{K}$ be a finite Kripke structure and $\rho$ and $\rho'$ be two tracks in $\mathrm{Trk}_{\mathcal{K}}$. We say that $\rho$ and $\rho'$ are *k-equivalent* if and only if, for every HS-formula $\psi$, with $\mathrm{Nest}_{BE}(\psi) = k$, $\mathcal{K}, \rho \models \psi$ if and only if $\mathcal{K}, \rho' \models \psi$.

It can be easily proved that *k*-equivalence propagates downwards.

**Proposition 1** *Let $\mathcal{K}$ be a finite Kripke structure and $\rho$ and $\rho'$ be two tracks in $\mathrm{Trk}_{\mathcal{K}}$. If $\rho$ and $\rho'$ are k-equivalent, then they are h-equivalent, for all $0 \leq h \leq k$.*

*Proof* Let us assume that $\mathcal{K}, \rho \models \psi$, with $0 \leq \mathrm{Nest}_{BE}(\psi) \leq k$. Consider the formula $\langle B \rangle^k \top$, whose BE-nesting depth is equal to $k$. It trivially holds that either $\mathcal{K}, \rho \models \langle B \rangle^k \top$ or $\mathcal{K}, \rho \models \neg \langle B \rangle^k \top$. In the first case, we have that $\mathcal{K}, \rho \models \langle B \rangle^k \top \wedge \psi$. Since $\mathrm{Nest}_{BE}\left(\langle B \rangle^k \top \wedge \psi\right) = k$, from the hypothesis, it immediately follows that $\mathcal{K}, \rho' \models \langle B \rangle^k \top \wedge \psi$, and thus $\mathcal{K}, \rho' \models \psi$. The other case can be dealt with in a symmetric way. □

We are now ready to introduce the notion of *descriptor*, which will play a fundamental role in the definition of finite abstract interval models.

**Definition 9** (*B*-descriptor and *E*-descriptor) Let $\mathcal{K} = (\mathcal{AP}, W, \delta, \mu, w_0)$ be a finite Kripke structure. A *B*-descriptor (resp., *E*-descriptor) is a labelled tree $\mathcal{D} = (\mathcal{V}, \mathcal{E}, \lambda)$, where $\mathcal{V}$ is a finite set of vertices, $\mathcal{E} \subseteq \mathcal{V} \times \mathcal{V}$ is a set of edges, and $\lambda : \mathcal{V} \mapsto W \times 2^W \times W$ is a node labelling function, that satisfies the following conditions:

1. for all $(v, v') \in \mathcal{E}$, with $\lambda(v) = (v_{in}, S, v_{fin})$ and $\lambda(v') = (v'_{in}, S', v'_{fin})$, it holds that $S' \subseteq S$, $v_{in} = v'_{in}$, and $v'_{fin} \in S$ (resp., $S' \subseteq S$, $v_{fin} = v'_{fin}$, and $v'_{in} \in S$);
2. for all pairs of edges $(v, v'), (v, v'') \in \mathcal{E}$, if the subtree rooted in $v'$ is isomorphic to the subtree rooted in $v''$, then $v' = v''$ (here and in the following, we write subtree for maximal subtree).

Condition (2) of Definition 9 simply states that no two subtrees, whose roots are siblings, can be isomorphic (notice that $\lambda$ is taken into account).

For $X \in \{B, E\}$, the *depth* of an *X*-descriptor $(\mathcal{V}, \mathcal{E}, \lambda)$ is the depth of the tree $(\mathcal{V}, \mathcal{E})$. We call an *X*-descriptor of depth $k \in \mathbb{N}$ an $X_k$-descriptor. An $X_0$-descriptor $\mathcal{D}$ consists of its root only, which is denoted by $\mathrm{root}(\mathcal{D})$. A label of a node will be referred to as a *descriptor element*. Hereafter, two descriptors will be considered *equal up to isomorphism*. The following proposition holds.

**Proposition 2** *For all $k \in \mathbb{N}$, there exists a finite number of possible $B_k$-descriptors (resp., $E_k$-descriptors).*

*Proof* Let us consider the case of $B_k$-descriptors (the case of $E_k$-descriptors is analogous). For $k = 0$, there are at most $|W| \cdot 2^{|W|} \cdot |W|$ pairwise distinct $B_0$-descriptors. As for the inductive step, let us assume $h$ to be the number of pairwise distinct *B*-descriptors of depth at most $k$. The number of $B_{k+1}$-descriptors is at most $|W| \cdot 2^{|W|} \cdot |W| \cdot 2^h$ (there are at most $|W| \cdot 2^{|W|} \cdot |W|$ possible choices for the root, which can have any subset of the $h$ $B$-descriptors of depth at most $k$ as subtrees). Moreover, by the König's lemma, they are all finite, because their depth is $k+1$ and the root has a finite number of children (no two subtrees of the root can be isomorphic). □

Proposition 2 provides an upper bound to the number of distinct $B_k$-descriptors (resp., $E_k$-descriptors), and thus to the number of nodes of each $B_{k+1}$-descriptor (resp., $E_{k+1}$-descriptors), for $k \in \mathbb{N}$, which is *not* elementary with respect to $|W|$ and $k$, $|W|$ being the exponent and $k$ the height of the exponential tower. As a matter of fact, this is a very rough upper bound, as some descriptors may not have depth $k+1$ and some of the "generated" trees might not even fulfil the definition of descriptor.

We show now how $B$-descriptors and $E$-descriptors can be exploited to extract relevant information from the tracks of a finite Kripke structure to be used in model checking. Let $\mathcal{K}$ be a finite Kripke structure and $\rho$ be a track in $\text{Trk}_\mathcal{K}$. For any $k \geq 0$, the label of the root of both the $B_k$-descriptor and $E_k$-descriptor for $\rho$ is the triple $(\text{fst}(\rho), \text{intstates}(\rho), \text{lst}(\rho))$. The root of the $B_k$-descriptor has a child for each prefix $\rho'$ of $\rho$, labelled with $(\text{fst}(\rho'), \text{intstates}(\rho'), \text{lst}(\rho'))$. Such a construction is then iteratively applied to the children of the root until either depth $k$ is reached or a track of length 2 is being considered on a node. The $E_k$-descriptor is built in a similar way by considering the suffixes of $\rho$.

In general, $B$- and $E$-descriptors do not convey enough information to determine which track they were built from (this will be clear shortly). However, they can be exploited to determine which HS formulas are satisfied by the track from which they have been built:

– to check satisfiability of proposition letters, they keep information about initial, final, and internal states of the track;
– to deal with $\langle A \rangle \psi$ and $\langle \overline{A} \rangle \psi$ formulas they store the final and initial states of the track;
– to deal with $\langle B \rangle \psi$ formulas, the $B$-descriptor keeps information about all the prefixes of the track;
– to deal with $\langle E \rangle \psi$ formulas, the $E$-descriptor keeps information about all the suffixes of the track;
– no additional information is needed for $\langle \overline{B} \rangle \psi$ and $\langle \overline{E} \rangle \psi$ formulas.

Let $\mathcal{K}$ be a finite Kripke structure. The $B_k$-descriptor (resp., $E_k$-descriptor) for a track $\rho$ in $\text{Trk}_\mathcal{K}$ is formally defined as follows.

**Definition 10** Let $\mathcal{K}$ be a finite Kripke structure, $\rho$ be a track in $\text{Trk}_\mathcal{K}$, and $k \in \mathbb{N}$. The $B_k$-descriptor (resp., $E_k$-descriptor) for $\rho$ is inductively defined as follows:

– for $k=0$, the $B_k$-descriptor (resp., $E_k$-descriptor) for $\rho$ is the tree $\mathcal{D} = (\text{root}(\mathcal{D}), \emptyset, \lambda)$, where
$$\lambda(\text{root}(\mathcal{D})) = (\text{fst}(\rho), \text{intstates}(\rho), \text{lst}(\rho));$$
– for $k > 0$, the $B_k$-descriptor (resp., $E_k$-descriptor) for $\rho$ is the tree $\mathcal{D} = (\mathcal{V}, \mathcal{E}, \lambda)$, where
$$\lambda(\text{root}(\mathcal{D})) = (\text{fst}(\rho), \text{intstates}(\rho), \text{lst}(\rho)),$$
which satisfies the following conditions:

1. for each prefix (resp., suffix) $\rho'$ of $\rho$, there exists $v \in \mathcal{V}$ such that $(\text{root}(\mathcal{D}), v) \in \mathcal{E}$ and the subtree rooted in $v$ is the $B_{k-1}$-descriptor (resp., $E_{k-1}$-descriptor) for $\rho'$;
2. for each vertex $v \in \mathcal{V}$ such that $(\text{root}(\mathcal{D}), v) \in \mathcal{E}$, there exists a prefix (resp., suffix) $\rho'$ of $\rho$ such that the subtree rooted in $v$ is the $B_{k-1}$-descriptor (resp., $E_{k-1}$-descriptor) for $\rho'$;
3. for all pairs of edges $(\text{root}(\mathcal{D}), v'), (\text{root}(\mathcal{D}), v'') \in \mathcal{E}$, if the subtree rooted in $v'$ is isomorphic to the subtree rooted in $v''$, then $v' = v''$.

It can be easily checked that any $B_k$-descriptor (resp., $E_k$-descriptor) for some track of some finite Kripke structure satisfies the conditions of Definition 9 (in particular, condition (1)), but not vice versa.

Consider, for instance, the $B_1$-descriptor reported in Figure 3. It is built on a set of states $W$ including at least states $v_0, v_1, v_2$, and $v_3$, and it satisfies both conditions of Definition 9. However, no track of a finite Kripke structure can be described by it, as no track may feature two prefixes to associate with the first two children of the root.

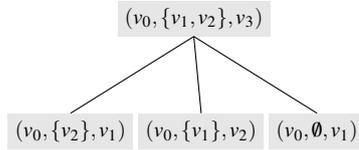

**Fig. 3** An example of a $B_1$-descriptor devoid of a corresponding track (in any Kripke structure).

*Example 3* In Figure 4, we depict the $B_2$- and $E_2$-descriptors for the track $v_0 v_1 v_0 v_0 v_1$ of the Kripke structure $\mathcal{K}_{Equiv}$ of Figure 1.

*Example 4* In Figure 5, we show the $B_2$-descriptor for the track $\rho = v_0 v_1 v_0 v_0 v_0 v_0 v_1$ of $\mathcal{K}_{Equiv}$. It is worth noticing that there exist two distinct prefixes of track $\rho$, that is, the tracks $\rho' = v_0 v_1 v_0 v_0 v_0 v_0$ and $\rho'' = v_0 v_1 v_0 v_0 v_0$, which have the same $B_1$-descriptor. Since, according to Definition 9, no tree can occur more than once as a subtree of the same node (in this example, the root), in the $B_2$-descriptor for $\rho$ prefixes $\rho'$ and $\rho''$ are represented by the same tree (the first subtree of the root on the left). In general, it holds that the root of a descriptor for a track with $h$ proper prefixes does not necessarily have $h$ children.

*Example 5* This example shows that not all of the $B_k$-descriptors that can be generated from the set of states of a given finite Kripke structure are $B_k$-descriptors for some track of that structure. (The same holds for $E_k$-descriptors.) Let us consider the finite Kripke structure $\mathcal{K}$ and the $B_1$-descriptor $\mathcal{D}_{B_1}$ respectively depicted on the left and the right of Figure 6. By inspecting $\mathcal{D}_{B_1}$, it can be easily checked that it can be the $B_1$-descriptor for tracks of the form $v_0 v_1^h v_3^2$, with $h \geq 2$, only. However, no track of this form can be obtained from the unravelling of $\mathcal{K}$.

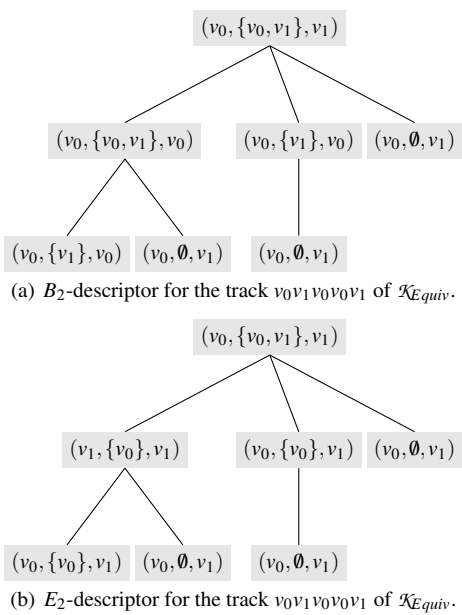

Fig. 4 $B_2$- and $E_2$-descriptors for the track $v_0v_1v_0v_0v_1$ of $\mathcal{K}_{Equiv}$.

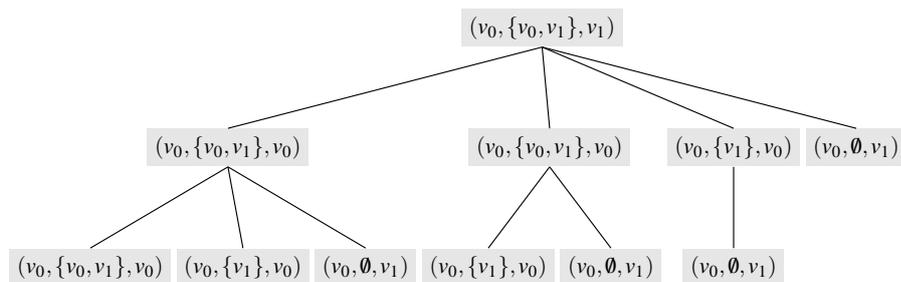

Fig. 5 The $B_2$-descriptor for the track $v_0v_1v_0v_0v_0v_0v_1$ of $\mathcal{K}_{Equiv}$.

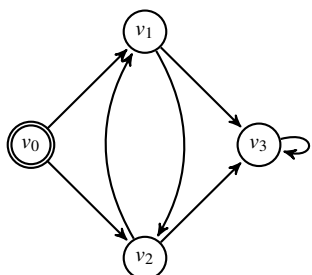

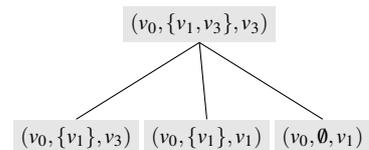

(a) A Kripke structure $\mathcal{K}$.

(b) $\mathcal{D}_{B_1}$: a $B_1$-descriptor not corresponding to any of the tracks of $\mathcal{K}$ in figure 6(a).

Fig. 6 Not all of the $B_k$-descriptors over $W$ are descriptors for some track of $\mathcal{K} = (\mathcal{AP}, W, \delta, \mu, w_0)$.

To check an HS formula against a given finite Kripke structure we actually need to account for both the *started-by* and *finished-by* relations at the same time. To this end, we introduce $BE_k$-descriptors for tracks. Given a finite Kripke structure $\mathcal{K}$ and a track $\rho$ in $\text{Trk}_{\mathcal{K}}$, the $BE_k$-descriptor for $\rho$ can be obtained from a suitable merging of its $B_k$-descriptor and $E_k$-descriptor. It can be viewed as a sort of "product" of the $B_k$-descriptor and the $E_k$-descriptor for $\rho$, and it is formally defined as follows:

**Definition 11** Let $\mathcal{K} = (\mathcal{AP}, W, \delta, \mu, w_0)$ be a finite Kripke structure, $\rho$ be a track in $\text{Trk}_{\mathcal{K}}$, and $k \in \mathbb{N}$. The $BE_k$-descriptor for $\rho$ is a labelled tree $\mathcal{D} = (\mathcal{V}, \mathcal{E}, \lambda)$, where $\mathcal{V}$ is a finite set of vertices, $\mathcal{E} = \mathcal{E}_B \cup \mathcal{E}_E$, with $\mathcal{E}_B \subseteq \mathcal{V} \times \mathcal{V}$ the set of "$B$-edges", $\mathcal{E}_E \subseteq \mathcal{V} \times \mathcal{V}$ the set of "$E$-edges", and $\mathcal{E}_B \cap \mathcal{E}_E = \emptyset$, and $\lambda : \mathcal{V} \mapsto W \times 2^W \times W$, which is inductively defined on $k \in \mathbb{N}$ as follows:

- for $k = 0$, the $BE_k$-descriptor for $\rho$ is $\mathcal{D} = (\text{root}(\mathcal{D}), \emptyset, \lambda)$, where

$$\lambda(\text{root}(\mathcal{D})) = (\text{fst}(\rho), \text{intstates}(\rho), \text{lst}(\rho)).$$

- for $k > 0$, the $BE_k$-descriptor for $\rho$ is $\mathcal{D} = (\mathcal{V}, \mathcal{E}, \lambda)$ with

$$\lambda(\text{root}(\mathcal{D})) = (\text{fst}(\rho), \text{intstates}(\rho), \text{lst}(\rho))$$

which satisfies the following conditions:
1a. for each prefix $\rho'$ of $\rho$, there exists $v \in \mathcal{V}$ such that $(\text{root}(\mathcal{D}), v) \in \mathcal{E}_B$ and the subtree rooted in $v$ is the $BE_{k-1}$-descriptor for $\rho'$;
1b. for each vertex $v \in \mathcal{V}$ such that $(\text{root}(\mathcal{D}), v) \in \mathcal{E}_B$, there exists a prefix $\rho'$ of $\rho$ such that the subtree rooted in $v$ is the $BE_{k-1}$-descriptor for $\rho'$;
1c. for all pairs of edges $(\text{root}(\mathcal{D}), v'), (\text{root}(\mathcal{D}), v'') \in \mathcal{E}_B$, if the subtree rooted in $v'$ is isomorphic to the subtree rooted in $v''$, then $v' = v''$;
2a. for each suffix $\rho''$ of $\rho$, there exists $v \in \mathcal{V}$ such that $(\text{root}(\mathcal{D}), v) \in \mathcal{E}_E$ and the subtree rooted in $v$ is the $BE_{k-1}$-descriptor for $\rho''$;
2b. for each vertex $v \in \mathcal{V}$ such that $(\text{root}(\mathcal{D}), v) \in \mathcal{E}_E$, there exists a suffix $\rho''$ of $\rho$ such that the subtree rooted in $v$ is the $BE_{k-1}$-descriptor for $\rho''$;
2c. for all pairs of edges $(\text{root}(\mathcal{D}), v'), (\text{root}(\mathcal{D}), v'') \in \mathcal{E}_E$, if the subtree rooted in $v'$ is isomorphic to the subtree rooted in $v''$, then $v' = v''$.

From Definition 11, it easily follows that for all $(v, v') \in \mathcal{E}_B$, with $\lambda(v) = (v_{in}, S, v_{fin})$ and $\lambda(v') = (v'_{in}, S', v'_{fin})$, $S' \subseteq S$, $v_{in} = v'_{in}$, and $v'_{fin} \in S$, and for all $(v, v') \in \mathcal{E}_E$, with $\lambda(v) = (v_{in}, S, v_{fin})$ and $\lambda(v') = (v'_{in}, S', v'_{fin})$, $S' \subseteq S$, $v_{fin} = v'_{fin}$ and $v'_{in} \in S$.

*Example 6* In Figure 7, with reference to the finite Krikpe structure $\mathcal{K}_{Equiv}$ of Figure 1, we give an example of a $BE_2$-descriptor. $B$-edges are represented by solid lines, while $E$-edges are represented by dashed lines. It is worth pointing out that the $BE_2$-descriptor of Figure 7 turns out to be the $BE_2$-descriptor for both the track $\rho = v_0 v_1 v_0^3 v_1$ and the track $\rho' = v_0 v_1 v_0^4 v_1$ (and many others). As we will see very soon, this is not an exception, but the rule: different tracks of a finite Kripke structures are described by the same $BE$-descriptor. Notice also that it features two isomorphic subtrees for the same node (the root). They both consist of a single node, labelled with $(v_0, \emptyset, v_1)$. However, this does not violate Definition 11 since one of them is connected to the parent via a $B$-edge and the other via an $E$-edge.

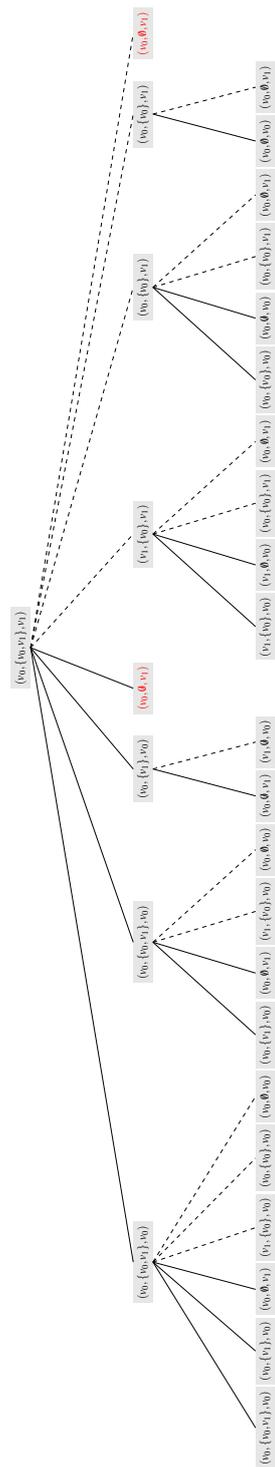

**Fig. 7** An example of $BE_2$-descriptor.

*Remark 1* It can be easily checked that the $BE_{k-1}$-descriptor $\mathcal{D}_{BE_{k-1}}$ for a track $\rho$ can be obtained from the $BE_k$-descriptor $\mathcal{D}_{BE_k}$ for such a track by removing the nodes at depth $k$ (if any) and the isomorphic subtrees possibly resulting from such a removal (see condition (1c) of Definition 11). In the following, we will sometimes denote $\mathcal{D}_{BE_{k-1}}$ by $\mathcal{D}_{BE_k}|_{k-1}$ to make it evident the way in which it is obtained.

$B_k$ and $E_k$-descriptors can be easily recovered from $BE_k$ ones. The $B_k$-descriptor $\mathcal{D}_{B_k}$ for a track $\rho$ can be obtained from the $BE_k$-descriptor $\mathcal{D}_{BE_k}$ for $\rho$ by pruning it in such a way that only those vertices of $\mathcal{D}_{BE_k}$ which are connected to the root via paths consisting of $B$-edges only are maintained (the set of edges of $\mathcal{D}_{B_k}$ and its labelling function can be obtained by restricting those of $\mathcal{D}_{BE_k}$ to the nodes of $\mathcal{D}_{B_k}$). The $E_k$-descriptor $\mathcal{D}_{E_k}$ of $\rho$ can be obtained in a similar way.

We focus now our attention on the relationships between the tracks obtained from the unravelling of a finite Kripke structure and their $BE_k$-descriptors. A key observation is that, even though the number of tracks of a finite Kripke structure $\mathcal{K}$ is infinite, for any $k \in \mathbb{N}$, the set of $BE_k$-descriptors for its tracks is finite. This is an immediate consequence of Definition 11 and Proposition 2. Thus, at least one $BE_k$-descriptor must be the $BE_k$-descriptor for infinitely many tracks. $BE_k$-descriptors naturally induce an equivalence relation of finite index over the set of tracks of a finite Kripke structure, that we call *k-descriptor equivalence relation*.

**Definition 12** Let $\mathcal{K}$ be a finite Kripke structure, $\rho, \rho'$ be two tracks in $\mathrm{Trk}_\mathcal{K}$, and $k \in \mathbb{N}$. We say that $\rho$ and $\rho'$ are *k-descriptor equivalent*, denoted by $\rho \sim_k \rho'$, if and only if the $BE_k$-descriptors for $\rho$ and $\rho'$ coincide.

The equivalence class of a track $\rho$ will be denoted by $[\rho]_{\sim_k}$. In the next section (Theorem 1), we will prove that, for any given pair of tracks $\rho, \rho' \in \mathrm{Trk}_\mathcal{K}$, if $\rho \sim_k \rho'$, then $\rho$ and $\rho'$ are $k$-equivalent (see Definition 8).

For all $k \in \mathbb{N}$, by exploiting the fact that the set of $BE_k$-descriptors for the tracks of a finite Kripke structure $\mathcal{K}$ is finite (or, equivalently, the equivalence relation $\sim_k$ has a finite index), we can associate a *finite* abstract interval model with $\mathcal{K}$, called the *quotient induced abstract interval model of depth k*, as follows.

Let $\mathcal{K}$ be a finite Kripke structure, $\mathrm{Trk}_\mathcal{K}$ be the set of all its tracks, and $k \in \mathbb{N}$. Each class of $\sim_k$ is identified by a $BE_k$-descriptor $\mathcal{D}_{BE_k}$, and it consists of all and only those tracks in $\mathrm{Trk}_\mathcal{K}$ which have $\mathcal{D}_{BE_k}$ as their $BE_k$-descriptor. We denote by $k$-Desc the set of all $BE_k$-descriptors $\mathcal{D}_{BE_k}$ such that *there exists at least one track $\rho$ in $\mathrm{Trk}_\mathcal{K}$ which is described by $\mathcal{D}_{BE_k}$* (we say that $\mathcal{D}_{BE_k}$ is witnessed by a track in $\mathrm{Trk}_\mathcal{K}$).

Allen's relations $A$ (*meets*), $B$ (*started-by*), and $E$ (*finished-by*) over $k$-Desc can be defined as follows.

**Definition 13** (Allen's relations $A$, $B$, $E$ over $k$-Desc) Let $\mathcal{D}_{BE_k}, \mathcal{D}'_{BE_k}$ be two $BE_k$-descriptors in $k$-Desc, with $\mathcal{D}_{BE_k} = (\mathcal{V}, \mathcal{E}_B \cup \mathcal{E}_E, \lambda)$ and $\mathcal{D}'_{BE_k} = (\mathcal{V}', \mathcal{E}'_B \cup \mathcal{E}'_E, \lambda')$. We say that:

1. $\left(\mathcal{D}_{BE_k}, \mathcal{D}'_{BE_k}\right) \in A_\mathrm{Desc}$ iff $\lambda\left(\mathrm{root}(\mathcal{D}_{BE_k})\right) = (v_{in}, S, v_{fin})$, $\lambda'\left(\mathrm{root}(\mathcal{D}'_{BE_k})\right) = (v'_{in}, S', v'_{fin})$, and $v_{fin} = v'_{in}$;

2. $\left(\mathcal{D}_{BE_k}, \mathcal{D}'_{BE_k}\right) \in B_{\text{Desc}}$ iff there exists $v \in \mathcal{V}$ such that $(\text{root}(\mathcal{D}_{BE_k}), v) \in \mathcal{E}_B$ and the subtree of $\mathcal{D}_{BE_k}$ rooted in $v$ is isomorphic to $\mathcal{D}'_{BE_k}|_{k-1}$;

3. $\left(\mathcal{D}_{BE_k}, \mathcal{D}'_{BE_k}\right) \in E_{\text{Desc}}$ iff there exists $v \in \mathcal{V}$ such that $(\text{root}(\mathcal{D}_{BE_k}), v) \in \mathcal{E}_E$ and the subtree of $\mathcal{D}_{BE_k}$ rooted in $v$ is isomorphic to $\mathcal{D}'_{BE_k}|_{k-1}$.

Definition 13 can be read as follows. Item 1 states that, whenever the third component (final state) of the label of the root of a $BE_k$-descriptor is equal to the first component (initial state) of the label of the root of another $BE_k$-descriptor, the two $BE_k$-descriptor are related by $A_{\text{Desc}}$. This amounts to say that any pair of tracks $\rho, \rho'$, which are described respectively by the former and latter $BE_k$-descriptor, are such that $\text{lst}(\rho) = \text{fst}(\rho')$, and thus Allen relation $A$ holds between $\rho$ and $\rho'$. Item 2 states that, whenever there exists a subtree of $\mathcal{D}_{BE_k}$, linked to the root via a $B$-edge, which is isomorphic to the tree obtained from $\mathcal{D}'_{BE_k}$ by removing the nodes at depth $k$ (if any) and the isomorphic subtrees possibly resulting from such a removal (this is the case, for instance, with subtrees of $\mathcal{D}'_{BE_k}$ that differ on the labels of nodes at depth $k$ only), $\mathcal{D}_{BE_k}$ and $\mathcal{D}'_{BE_k}$ are related by $B_{\text{Desc}}$. As matter of fact, several tracks may be described by the same $BE_k$-descriptor $\mathcal{D}_{BE_k}$. However, whenever a track is described by (the tree obtained from the pruning of) $\mathcal{D}'_{BE_k}$, it is a prefix of at least one of the tracks described by $\mathcal{D}_{BE_k}$. Item 3 is analogous to item 2.

The generalisation of Definition 13 to pairs of descriptors belonging to $k$-Desc and $k'$-Desc, with $k \neq k'$, is straightforward.

We are now ready to formally define the notion of quotient induced abstract interval model of depth $k$.

**Definition 14** (Quotient induced abstract interval model of depth $k$) Let $\mathcal{K} = (\mathcal{AP}, W, \delta, \mu, v_0)$ be a finite Kripke structure, $\varphi$ be an HS formula with BE-nesting depth $k \in \mathbb{N}$, and

$$\Omega = \bigcup_{h \leq k} h\text{-Desc}.$$

The quotient induced abstract interval model of depth $k$ is the *finite* abstract interval model $\mathcal{A}/\sim_k = (\mathcal{AP}, \Omega, A_{\text{Desc}}, B_{\text{Desc}}, E_{\text{Desc}}, \sigma)$, where the valuation function $\sigma : \Omega \mapsto 2^{\mathcal{AP}}$ is such that for all $\mathcal{D}_{BE} \in \Omega$, with $\lambda(\text{root}(\mathcal{D}_{BE})) = (v_{in}, S, v_{fin})$,

$$\sigma(\mathcal{D}_{BE}) = \mu(v_{in}) \cap \bigcap_{v \in S} \mu(v) \cap \mu(v_{fin}).$$

## 4 Decidability of model checking for HS over finite Kripke structures

In this section, we prove the decidability of the model checking problem for HS over finite Kripke structures (under the homogeneity assumption). The proof makes an essential use of quotient induced abstract interval models. Formally, we show that, for any given finite Kripke structure $\mathcal{K}$, the (finite) quotient induced abstract interval model $\mathcal{A}/\sim_k$ and the (infinite) abstract interval model $\mathcal{A}_\mathcal{K}$, induced by $\mathcal{K}$, are equivalent with respect to the satisfiability of HS formulas with nesting depth at most $k$. In addition, we show that the notions of $k$-equivalence and $k$-descriptor equivalence are not equivalent (if two tracks are *k*-descriptor equivalent, they are also *k*-equivalent, but not vice versa), and we show how to weaken the notion of *k*-descriptor equivalence to perfectly match *k*-equivalence.

4.1 The decidability proof

As a preliminary step, we prove a right extension property. Let $\mathcal{K}$ be a finite Kripke structure, $k \in \mathbb{N}$, and $\rho$ and $\rho'$ be two tracks in $\text{Trk}_{\mathcal{K}}$ with the same $BE_k$-descriptor (and thus, in particular, $\text{lst}(\rho) = \text{lst}(\rho')$). The property states that if we extend $\rho$ and $\rho'$ "to the right" with the same track $\overline{\rho}$ in $\text{Trk}_{\mathcal{K}}$, with $(\text{lst}(\rho), \text{fst}(\overline{\rho})) \in \delta$, then the resulting tracks $\rho \cdot \overline{\rho}$ and $\rho' \cdot \overline{\rho}$ (both belonging to $\text{Trk}_{\mathcal{K}}$) have the same $BE_k$-descriptor as well. An analogous property holds for the extension of the two tracks $\rho$ and $\rho'$ "to the left", which guarantees that $\overline{\rho} \cdot \rho$ and $\overline{\rho} \cdot \rho'$ have the same $BE_k$-descriptor (left extension property). In the proof, we will exploit the fact that if two tracks in $\text{Trk}_{\mathcal{K}}$ have the same $BE_{k+1}$-descriptor, then they also have the same $BE_k$-descriptor (see Remark 1).

**Proposition 3** *(Right extension property) Let $\mathcal{K} = (\mathcal{AP}, W, \delta, \mu, v_0)$ be a finite Kripke structure and let $\rho$ and $\rho'$ be two tracks in $\text{Trk}_{\mathcal{K}}$ with the same $BE_k$-descriptor. For any track $\overline{\rho}$ in $\text{Trk}_{\mathcal{K}}$, with $(\text{lst}(\rho), \text{fst}(\overline{\rho})) \in \delta$, the two tracks $\rho \cdot \overline{\rho}$ and $\rho' \cdot \overline{\rho}$ belong to $\text{Trk}_{\mathcal{K}}$ and have the same $BE_k$-descriptor.*

*Proof* The proof is by induction on $k \in \mathbb{N}$.

- Base case ($k = 0$). Since $\rho$ and $\rho'$ have the same $BE_0$-descriptor, it holds that $\text{fst}(\rho) = \text{fst}(\rho'), \text{intstates}(\rho) = \text{intstates}(\rho')$, and $\text{lst}(\rho) = \text{lst}(\rho')$ and thus
  - $\text{fst}(\rho \cdot \overline{\rho}) = \text{fst}(\rho) = \text{fst}(\rho') = \text{fst}(\rho' \cdot \overline{\rho})$;
  - $\text{lst}(\rho \cdot \overline{\rho}) = \text{lst}(\rho' \cdot \overline{\rho}) = \text{lst}(\overline{\rho})$;
  - $\text{intstates}(\rho \cdot \overline{\rho}) = \text{intstates}(\rho) \cup \{\text{lst}(\rho), \text{fst}(\overline{\rho})\} \cup \text{intstates}(\overline{\rho}) = \text{intstates}(\rho') \cup \{\text{lst}(\rho'), \text{fst}(\overline{\rho})\} \cup \text{intstates}(\overline{\rho}) = \text{intstates}(\rho' \cdot \overline{\rho})$

  This allows us to conclude that $\rho \cdot \overline{\rho}$ and $\rho' \cdot \overline{\rho}$ have the same $BE_0$-descriptor.
- Inductive step ($k > 0$). Let $\overline{\mathcal{D}_{BE_k}} = (\overline{\mathcal{V}}, \overline{\mathcal{E}_B} \cup \overline{\mathcal{E}_E}, \overline{\lambda})$ and $\overline{\mathcal{D}_{BE_k}}' = (\overline{\mathcal{V}}', \overline{\mathcal{E}_B}' \cup \overline{\mathcal{E}_E}', \overline{\lambda}')$ be respectively the $BE_k$-descriptors of $\rho \cdot \overline{\rho}$ and $\rho' \cdot \overline{\rho}$. We prove that $\overline{\mathcal{D}_{BE_k}}$ and $\overline{\mathcal{D}_{BE_k}}'$ are equal (up to isomorphism).

  As for the roots, the same argument we used for the base case can be exploited to prove that $\overline{\lambda}(\text{root}(\overline{\mathcal{D}_{BE_k}})) = \overline{\lambda}'(\text{root}(\overline{\mathcal{D}_{BE_k}}'))$ (they have the same labelling).
  Let us consider now a node $v \in \overline{\mathcal{V}}$ such that $(\text{root}(\overline{\mathcal{D}_{BE_k}}), v) \in \overline{\mathcal{E}_B}$ (resp., $\overline{\mathcal{E}_E}$). We show that there exists a node $v' \in \overline{\mathcal{V}}'$ such that $(\text{root}(\overline{\mathcal{D}_{BE_k}}'), v') \in \overline{\mathcal{E}_B}'$ (resp., $\overline{\mathcal{E}_E}'$) and the subtrees rooted in $v$ and in $v'$ are isomorphic.
  - Let us consider the case $(\text{root}(\overline{\mathcal{D}_{BE_k}}), v) \in \overline{\mathcal{E}_B}$. By definition of $BE_k$-descriptor, there exists a prefix $\rho''$ of $\rho \cdot \overline{\rho}$ such that the subtree rooted in $v$ is the $BE_{k-1}$-descriptor of $\rho''$. Three cases are possible.
    * Case 1: $\rho''$ is a (proper) prefix of $\rho$. Since $\rho$ and $\rho'$ have the same $BE_k$-descriptor, there is a prefix $\rho'''$ of $\rho'$ with the same $BE_{k-1}$-descriptor as $\rho''$.

- Case 2: $\rho'' = \rho$. Since $\rho$ and $\rho'$ have the same $BE_k$-descriptor, they have also the same $BE_{k-1}$-descriptor (see Remark 1).
- Case 3: $\rho'' = \rho \cdot \tilde{\rho}$, where $\tilde{\rho}$ is a prefix of $\overline{\rho}$. As pointed out in Remark 1, we know that $\rho$ and $\rho'$ have the same $BE_h$-descriptor, for all $h \leq k$. Then, by the inductive hypothesis, $\rho \cdot \tilde{\rho}$ and $\rho' \cdot \tilde{\rho}$ have the same $BE_{k-1}$-descriptor.

In all three cases, by Definition 11, we can conclude that there exists a node $v' \in \overline{\mathcal{V}}'$ such that $(\text{root}(\overline{\mathcal{D}_{BE_k}}'), v') \in \overline{\mathscr{E}_B}'$ and the subtrees rooted in $v$ and in $v'$ are isomorphic.

– Now, let $(\text{root}(\overline{\mathcal{D}_{BE_k}}), v) \in \overline{\mathscr{E}_E}$. By definition of $BE_k$-descriptor, there exists a suffix $\rho''$ of $\rho \cdot \overline{\rho}$ such that the subtree rooted in $v$ is the $BE_{k-1}$-descriptor of $\rho''$. We distinguish two cases.
- Let $\rho''$ be a proper suffix of $\overline{\rho}$ or $\rho'' = \overline{\rho}$. Then, $\rho''$ is a suffix of both $\rho \cdot \overline{\rho}$ and $\rho' \cdot \overline{\rho}$. Hence, the same $BE_{k-1}$-descriptor is rooted both in $v$ and in $v'$, for some $v' \in \overline{V}'$ such that $(\text{root}(\overline{\mathcal{D}_{BE_k}}'), v') \in \overline{\mathscr{E}_E}'$.
- Let $\rho'' = \tilde{\rho} \cdot \overline{\rho}$, where $\tilde{\rho}$ is a suffix of $\rho$. If $|\tilde{\rho}| = 1$, $\rho''$ is a suffix of both $\rho \cdot \overline{\rho}$ and $\rho' \cdot \overline{\rho}$, as $\text{lst}(\rho) = \text{lst}(\rho')$. Let $|\tilde{\rho}| \geq 2$. Since by hypothesis $\rho$ and $\rho'$ have the same $BE_k$-descriptor, there is a subtree of depth $k-1$ in this descriptor which is associated both with $\tilde{\rho}$ and with a suffix of $\rho'$, say, $\tilde{\rho}'$. By inductive hypothesis, $\rho'' = \tilde{\rho} \cdot \overline{\rho}$ and $\tilde{\rho}' \cdot \overline{\rho}$ have the same $BE_{k-1}$-descriptor. In both cases ($|\tilde{\rho}| = 1$ and $|\tilde{\rho}| \geq 2$), it immediately follows that there exists a node $v' \in \overline{\mathcal{V}}'$, which is the root of the subtree for $\text{lst}(\rho') \cdot \overline{\rho}$ (resp., $\tilde{\rho}' \cdot \overline{\rho}$), such that $(\text{root}(\overline{\mathcal{D}_{BE_k}}'), v') \in \overline{\mathscr{E}_E}'$ and the subtrees rooted in $v$ and in $v'$ are isomorphic.

To sum up, we have shown that (i) $\overline{\lambda}(\text{root}(\overline{\mathcal{D}_{BE_k}})) = \overline{\lambda}'(\text{root}(\overline{\mathcal{D}_{BE_k}}'))$, (ii) for each prefix of $\rho \cdot \overline{\rho}$ there exists a prefix of $\rho' \cdot \overline{\rho}$ with the same $BE_{k-1}$-descriptor, and (iii) for each suffix of $\rho \cdot \overline{\rho}$ there exists a suffix of $\rho' \cdot \overline{\rho}$ with the same $BE_{k-1}$-descriptor. The converse of conditions (ii) and (iii) holds by symmetry. This allows us to conclude that $\overline{\mathcal{D}_{BE_k}}$ and $\overline{\mathcal{D}_{BE_k}}'$ are isomorphic.
□

The next theorem proves that $k$-descriptor equivalent tracks are $k$-equivalent.

**Theorem 1** *($k$-descriptor equivalence implies $k$-equivalence) Let $\psi$ be an HS formula, with $\text{Nest}_{BE}(\psi) = k$, $\mathcal{K}$ be a finite Kripke structure, $\rho$ and $\rho'$ be two tracks in $\text{Trk}_{\mathcal{K}}$, and $\mathcal{A}_{\mathcal{K}}$ be the abstract interval model induced by $\mathcal{K}$. If $\rho$ and $\rho'$ have the same $BE_k$-descriptor, then*

$$\mathcal{A}_{\mathcal{K}}, \rho \models \psi \iff \mathcal{A}_{\mathcal{K}}, \rho' \models \psi$$

*Proof* The proof is by induction on the structural complexity of $\psi$.

– $\psi = p$: $\mathcal{A}_{\mathcal{K}}, \rho \models p$ iff $p \in \bigcap_{w \in \text{states}(\rho)} \mu(w)$. Since $\rho$ and $\rho'$ have the same $BE_k$-descriptor, they consist of occurrences of the same set of states of $\mathcal{K}$, that is, $\text{states}(\rho) = \text{states}(\rho')$, witnessed by the root of the $BE_k$-descriptor. Therefore, $\mathcal{A}_{\mathcal{K}}, \rho \models p$ iff $\mathcal{A}_{\mathcal{K}}, \rho' \models p$.
– $\psi = \neg \varphi$: $\mathcal{A}_{\mathcal{K}}, \rho \models \psi$ iff $\mathcal{A}_{\mathcal{K}}, \rho \not\models \varphi$ iff (by inductive hypothesis) $\mathcal{A}_{\mathcal{K}}, \rho' \not\models \varphi$ iff $\mathcal{A}_{\mathcal{K}}, \rho' \models \psi$.

- $\psi = \varphi_1 \wedge \varphi_2$: let us assume that $\text{Nest}_{BE}(\varphi_1) = \text{Nest}_{BE}(\psi) = k$ and $\text{Nest}_{BE}(\varphi_2) \leq k$. By the inductive hypothesis, $\mathcal{A}_{\mathcal{K}}, \rho \models \varphi_1$ iff $\mathcal{A}_{\mathcal{K}}, \rho' \models \varphi_1$. Since any pair of tracks that have the same $BE_k$-descriptor have also the same $BE_{k'}$-descriptor, for all $k' \leq k$ (see Remark 1), by the inductive hypothesis, $\mathcal{A}_{\mathcal{K}}, \rho \models \varphi_2$ iff $\mathcal{A}_{\mathcal{K}}, \rho' \models \varphi_2$. Hence, if $\mathcal{A}_{\mathcal{K}}, \rho \models \psi$, then $\mathcal{A}_{\mathcal{K}}, \rho \models \varphi_1$ and $\mathcal{A}_{\mathcal{K}}, \rho \models \varphi_2$, and thus $\mathcal{A}_{\mathcal{K}}, \rho' \models \psi$. As for the converse, if $\mathcal{A}_{\mathcal{K}}, \rho' \models \psi$, then $\mathcal{A}_{\mathcal{K}}, \rho' \models \varphi_1$ and $\mathcal{A}_{\mathcal{K}}, \rho' \models \varphi_2$, and thus $\mathcal{A}_{\mathcal{K}}, \rho \models \psi$.
- $\psi = \langle A \rangle \varphi$: $\mathcal{A}_{\mathcal{K}}, \rho \models \psi$ iff there exists $\overline{\rho} \in \text{Trk}_{\mathcal{K}}$ such that $\text{lst}(\rho) = \text{fst}(\overline{\rho})$ and $\mathcal{A}_{\mathcal{K}}, \overline{\rho} \models \varphi$. Analogously, $\mathcal{A}_{\mathcal{K}}, \rho' \models \psi$ iff there exists $\overline{\rho}' \in \text{Trk}_{\mathcal{K}}$ such that $\text{lst}(\rho') = \text{fst}(\overline{\rho}')$ and $\mathcal{A}_{\mathcal{K}}, \overline{\rho}' \models \varphi$. Since $\rho$ and $\rho'$ have the same $BE_k$-descriptor, it holds that $\text{lst}(\rho) = \text{lst}(\rho')$. Hence, we can choose $\overline{\rho} = \overline{\rho}'$, so that $\mathcal{A}_{\mathcal{K}}, \overline{\rho} \models \varphi$ iff $\mathcal{A}_{\mathcal{K}}, \overline{\rho}' \models \varphi$.
- $\psi = \langle \overline{A} \rangle \varphi$: analogous to the previous case.
- $\psi = \langle B \rangle \varphi$: $\text{Nest}_{BE}(\psi) = 1 + \text{Nest}_{BE}(\varphi) = k$. If $\mathcal{A}_{\mathcal{K}}, \rho \models \psi$, then there exists $\overline{\rho} \in \text{Pref}(\rho)$ such that $\mathcal{A}_{\mathcal{K}}, \overline{\rho} \models \varphi$. Let $\mathcal{D}_{BE_k} = (\mathcal{V}, \mathcal{E}_B \cup \mathcal{E}_E, \lambda)$ be the $BE_k$-descriptor for $\rho$. By definition of $BE_k$-descriptor, there exists an edge $(\text{root}(\mathcal{D}_{BE_k}), v) \in \mathcal{E}_B$ such that the subtree rooted in $v$ is the $BE_{k-1}$-descriptor for $\overline{\rho}$. Since, by hypothesis, $\rho$ and $\rho'$ have the same $BE_k$-descriptor, there exists a prefix $\overline{\rho}'$ of $\rho'$ such that the subtree rooted in $v$ is the $BE_{k-1}$-descriptor for $\overline{\rho}'$. Now, by the inductive hypothesis, $\mathcal{A}_{\mathcal{K}}, \overline{\rho}' \models \varphi$, and thus $\mathcal{A}_{\mathcal{K}}, \rho' \models \psi$. Exactly the same argument allows us to conclude that if $\mathcal{A}_{\mathcal{K}}, \rho' \models \psi$, then $\mathcal{A}_{\mathcal{K}}, \rho \models \psi$.
- $\psi = \langle \overline{B} \rangle \varphi$: if $\mathcal{A}_{\mathcal{K}}, \rho \models \psi$, then there exists $\overline{\rho}$ in $\text{Trk}_{\mathcal{K}}$ such that $\rho \in \text{Pref}(\overline{\rho})$ and $\mathcal{A}_{\mathcal{K}}, \overline{\rho} \models \varphi$. We can express $\overline{\rho}$ as $\rho \cdot \tilde{\rho}$ for some $\tilde{\rho}$ in $\text{Trk}_{\mathcal{K}}$ such that $(\text{lst}(\rho), \text{fst}(\tilde{\rho})) \in \delta$. Now, since $\rho$ and $\rho'$ have the same $BE_k$-descriptor, it holds that $\text{lst}(\rho) = \text{lst}(\rho')$. By Proposition 3, the tracks $\overline{\rho} = \rho \cdot \tilde{\rho}$ and $\rho' \cdot \tilde{\rho}$ have the same $BE_k$-descriptor. By the inductive hypothesis, $\mathcal{A}_{\mathcal{K}}, \rho' \cdot \tilde{\rho} \models \varphi$, and thus $\mathcal{A}_{\mathcal{K}}, \rho' \models \psi$. Exactly the same argument allows us to conclude that if $\mathcal{A}_{\mathcal{K}}, \rho' \models \psi$, then $\mathcal{A}_{\mathcal{K}}, \rho \models \psi$.
- $\psi = \langle E \rangle \varphi$ and $\psi = \langle \overline{E} \rangle \varphi$ are symmetric to $\psi = \langle B \rangle \varphi$ and $\psi = \langle \overline{B} \rangle \varphi$, respectively.

□

Since $k$-descriptor equivalence preserves satisfiability of HS formulas, testing whether $\mathcal{K}, \rho \models \psi$ can be reduced to checking whether $\mathcal{A}/\sim_k, [\rho]_{\sim_k} \models \psi$.

**Corollary 1** *Let $\psi$ be an HS formula, with $\text{Nest}_{BE}(\psi) \leq k$, $\mathcal{K}$ be a finite Kripke structure, and $\rho$ be a track in $\text{Trk}_{\mathcal{K}}$. It holds that*

$$\mathcal{K}, \rho \models \psi \iff \mathcal{A}/\sim_k, [\rho]_{\sim_k} \models \psi.$$

*Proof* By Definition 5, $\mathcal{K}, \rho \models \psi$ if and only if $\mathcal{A}_{\mathcal{K}}, \rho \models \psi$. The proof of the left-to-right implication (if $\mathcal{A}_{\mathcal{K}}, \rho \models \psi$, then $\mathcal{A}/\sim_k, [\rho]_{\sim_k} \models \psi$) is by induction on the structural complexity of $\psi$, and it basically makes use of Definition 13 and Definition 14. The proof of the opposite implication is straightforward. □

By exploiting Corollary 1, we can reduce the model checking problem for HS against finite Kripke structures to the model checking problem for multi-modal, finite Kripke structures, whose nodes are all possible (witnessed) descriptors, with depth up to $k$, and there is a distinct accessibility relation for each one of the HS modalities $A$, $B$, $E$, $\overline{A}$, $\overline{B}$, and $\overline{E}$. Since the model checking problem for multi-modal, finite Kripke structures and formulas is decidable (in [9,13], it has been shown that the model checking problem for multi-modal Kripke structures and formulas is decidable in polynomial time with respect to both the size of the Kripke structure and the length of the formula), decidability of the model checking problem for HS against finite Kripke structures immediately follows.

**Theorem 2** *The model checking problem for HS against finite Kripke structures is decidable (with a* non-elementary *algorithm).*

*Proof* Let $\mathcal{K}$ be a finite Kripke structure and let $\varphi$ be the HS formula to check, with $\text{Nest}_{BE}(\varphi) = k$. We first prove that, in order to select the $BE_h$-descriptors, with $0 \leq h \leq k$, witnessed by some track in $\mathcal{K}$, we can restrict ourselves to tracks devoid of prefixes associated with the same $BE_k$-descriptor.

Let $\rho \in \text{Trk}_\mathcal{K}$ and let $\rho', \rho''$ be two prefixes of $\rho$, with $|\rho''| < |\rho'| \leq |\rho|$ (notice that we allow $\rho'$ to coincide with $\rho$). Moreover, let $\rho = \rho' \cdot \tilde{\rho}$, for some $\tilde{\rho}$ with $|\tilde{\rho}| \geq 1$ (in case $|\rho| = |\rho'|$, $\rho = \rho'$). If the $BE_k$-descriptors for $\rho'$ and $\rho''$ are the same, then, by Proposition 3, it holds that the $BE_k$-descriptor for $\rho'' \cdot \tilde{\rho}$ is equal to the one for $\rho' \cdot \tilde{\rho} = \rho$. Hence, we can safely replace $\rho$ by the $k$-descriptor equivalent shorter track $\rho'' \cdot \tilde{\rho}$. We can iterate such a contraction process until there are no more pairs of prefixes with the same $BE_k$-descriptor[1].

Proposition 2 provides a non-elementary upper bound $\alpha$ to the number of distinct $BE_h$-descriptors, with $0 \leq h \leq k$ (as well as to their size), with respect to the size of $\mathcal{K}$ and the nesting depth $k$ of $\varphi$. A bound on the length of the tracks in $\text{Trk}_\mathcal{K}$ that we need to consider in order to determine the witnessed $BE_h$-descriptors in an effective way immediately follows (it is equal to $1 + \alpha$, where 1 must be added because the length of any track is greater than or equal to 2).

Hence, in order to generate all the witnessed $BE_h$-descriptors, with $0 \leq h \leq k$, it suffices to list, for all states $v$ of $\mathcal{K}$, all the tracks starting from $v$, ordered by length, until the above bound is reached, and then to build the corresponding $BE_h$-descriptors, with $0 \leq h \leq k$.

This allows us to conclude that the derived model checking problem for multi-modal, finite Kripke structures has to be solved over a model whose size has a non-elementary upper bound. □

## 4.2 $k$-equivalence and corresponding $BE_k$-descriptors

In the previous section (Theorem 1), we proved that $k$-descriptor equivalence is a sufficient condition for $k$-equivalence, that is, if two tracks are $k$-descriptor equivalent, then they are $k$-equivalent. However, it is not a necessary one. To show that the converse does not hold, consider once more the finite Kripke structure $\mathcal{K}_{Equiv}$ in Figure

---

[1] As a matter of fact, the same argument can be given by referring to suffixes instead of prefixes. Anyway, as one can easily see, making use of both the right extension and the left extension properties does not allow us to improve the claimed bound.

1. The tracks $v_0^5$ and $v_0^6$ of $\mathcal{K}_{Equiv}$ have the same $BE_2$-descriptor, but not the same $BE_3$-descriptor, yet there exists no formula $\psi$, with $\text{Nest}_{BE}(\psi) \leq 3$, such that $\mathcal{K}, v_0^6 \models \psi$ and $\mathcal{K}, v_0^5 \not\models \psi$. Intuitively, since these two tracks are made of a different number of occurrences of the same state, the only way to distinguish them is by means of the formula $\langle B \rangle^4 \top$, or similar ones, for which $\mathcal{K}, v_0^6 \models \langle B \rangle^4 \top$ and $\mathcal{K}, v_0^5 \not\models \langle B \rangle^4 \top$, but these formulas have a BE-nesting depth *higher* than 3.

In the following, we introduce the notion of *corresponding $BE_k$-descriptors*, and we prove that it provides a necessary and sufficient condition for $k$-equivalence. Such a notion allows us to rephrase equivalence between tracks in terms of more abstract characteristics of their descriptors, in a stronger way than Theorem 1. As an example, by exploiting the correspondence among descriptors it defines and the statement of Theorem 3 below, it will be possible to prove that $v_0^5$ and $v_0^6$ are actually 3-equivalent.

We start by providing some definitions.

**Definition 15** Let $\mathcal{K} = (\mathcal{AP}, W, \delta, \mu, w_0)$ be a finite Kripke structure, $\mathcal{D}_{BE_k}$ be a $BE_k$-descriptor associated with a track of $\mathcal{K}$, and $(v_{in}, S, v_{fin})$ be the label of the root of $\mathcal{D}_{BE_k}$.

- Let $\rho$ be a track of $\mathcal{K}$ with $\text{fst}(\rho) = v_{fin}$. We say that the $BE_k$-descriptor for $\rho$ is an $A$-successors of $\mathcal{D}_{BE_k}$.
- Let $\tilde{\rho}$ be a track of $\mathcal{K}$ associated with $\mathcal{D}_{BE_k}$ and $\rho$ be a track of $\mathcal{K}$ with $(v_{fin}, \text{fst}(\rho)) \in \delta$. We say that the $BE_k$-descriptor for $\tilde{\rho} \cdot \rho$ is a $\overline{B}$-successor of $\mathcal{D}_{BE_k}$[2].

The definitions of $\overline{A}$-successors and $\overline{E}$-successors can easily be obtained by symmetry. Since, in a finite Kripke structure, every state has (at least) a successor with respect to $\delta$, $BE_k$-descriptors always have both $A$-successors and $\overline{B}$-successors. On the contrary, $BE_k$-descriptors may have no $\overline{A}$-successors or $\overline{E}$-successors, because a state does not necessarily have a predecessor with respect to $\delta$.

The set of descriptors witnessed by some tracks of $\mathcal{K}$ and their successor relations, corresponding to the various HS modalities, allow us to define a graph structure.

**Definition 16** (Graph $\mathcal{G}$ of the $BE$-descriptors for the tracks of $\mathcal{K}$) Let $\mathcal{K} = (\mathcal{AP}, W, \delta, \mu, w_0)$ be a finite Kripke structure. The graph $\mathcal{G}$ of the $BE$-descriptors of depth at most $k$, with $k \geq 0$, witnessed by some tracks of $\mathcal{K}$, is a pair $(\mathcal{V}_\mathcal{G}, \mathcal{E}_\mathcal{G})$, where $\mathcal{E}_\mathcal{G} \subseteq \mathcal{V}_\mathcal{G} \times \mathcal{V}_\mathcal{G}$ is a set of labelled edges, such that:

- $\mathcal{V}_\mathcal{G}$ contains a node for each $BE_h$-descriptor, with $0 \leq h \leq k$, witnessed by some track of $\mathcal{K}$;
- edges in $\mathcal{E}_\mathcal{G}$ are labelled with $X \in \{A, B, E, \overline{A}, \overline{B}, \overline{E}\}$ according to the following criteria:
  - $(v, v') \in \mathcal{E}_\mathcal{G}$ is an $X$-edge, with $X \in \{A, \overline{A}, \overline{B}, \overline{E}\}$, whenever the descriptor of $v'$ is an $X$-successor of the descriptor of $v$;
  - $(v, v') \in \mathcal{E}_\mathcal{G}$ is a $B$-edge whenever the descriptors associated with $v$ and $v'$ are $\mathcal{D}_{BE_h}$ and $\mathcal{D}'_{BE_{h-1}}$, respectively (for some $h \geq 1$), and $\mathcal{D}'_{BE_{h-1}}$ is isomorphic to a subtree of $\mathcal{D}_{BE_h}$ connected to the root of $\mathcal{D}_{BE_h}$ via a $B$-edge;

---
[2] If a track $\overline{\rho}$ was considered in place of $\tilde{\rho}$, with the same $BE_k$-descriptor $\mathcal{D}_{BE_k}$ as $\tilde{\rho}$, by the right extension property, both $\tilde{\rho} \cdot \rho$ and $\overline{\rho} \cdot \rho$ are associated with the same descriptor as well.

- $(v, v') \in \mathscr{E}_\mathscr{G}$ is an *E*-edge whenever the descriptors associated with $v$ and $v'$ are $\mathcal{D}_{BE_h}$ and $\mathcal{D}'_{BE_{h-1}}$, respectively (for some $h \geq 1$), and $\mathcal{D}'_{BE_{h-1}}$ is isomorphic to a subtree of $\mathcal{D}_{BE_h}$ connected to the root of $\mathcal{D}_{BE_h}$ via an *E*-edge.

The set of nodes $\mathscr{V}_\mathscr{G}$ is finite and the out-degree of every node is finite as well. Moreover, $\mathscr{V}_\mathscr{G}$ can be partitioned into $k$ sets, according to the depth of the descriptors associated with its nodes. A node associated with a descriptor of depth $h$ can be connected to a node associated with a descriptor of depth $h-1$, with $0 < h \leq k$, only by *B*- or *E*-edges. The number of proper prefixes (resp., suffixes) of short enough tracks can indeed be less than $k$. In such a case, the actual height of $BE_k$-descriptors is less than the nominal height $k$, and thus it may happen that the $BE_i$-descriptor and $BE_j$-descriptor, with $i \neq j$, for a track are isomorphic. When collecting all the $BE_i$-descriptor, for $0 \leq i \leq k$, isomorphic descriptors of different depths will be considered as distinct nodes of $\mathscr{V}_\mathscr{G}$.

The notion of *corresponding $BE_k$-descriptors up to depth $n$* is defined as follows.

**Definition 17** Let $\mathcal{K} = (\mathcal{AP}, W, \delta, \mu, w_0)$ be a finite Kripke structure, $\mathcal{D}_{BE_k}$ and $\mathcal{D}'_{BE_k}$ be two $BE_k$-descriptors associated with some of its tracks, and $(v_{in}, S, v_{fin})$ and $(v'_{in}, S', v'_{fin})$ be the labels of the root of $\mathcal{D}_{BE_k}$ and $\mathcal{D}'_{BE_k}$, respectively. We say that $\mathcal{D}_{BE_k}$ and $\mathcal{D}'_{BE_k}$ are *corresponding $BE_k$-descriptors up to depth $n$* if and only if:

- the two roots are labelled by the same set of propositions, that is,

$$\bigcap_{w \in \{v_{in}\} \cup S \cup \{v_{fin}\}} \mu(w) = \bigcap_{w' \in \{v'_{in}\} \cup S' \cup \{v'_{fin}\}} \mu(w');$$

- if $n > 0$:
  - for any track $\rho \in \text{Trk}_\mathcal{K}$, with $\text{fst}(\rho) = v_{fin}$, there is a track $\rho' \in \text{Trk}_\mathcal{K}$, with $\text{fst}(\rho') = v'_{fin}$, such that $\rho$ and $\rho'$ are associated with corresponding $BE_k$-descriptors up to depth $n-1$, and vice versa;
  - for any track $\rho \in \text{Trk}_\mathcal{K}$, with $\text{lst}(\rho) = v_{in}$, there is a track $\rho' \in \text{Trk}_\mathcal{K}$, with $\text{lst}(\rho') = v'_{in}$, such that $\rho$ and $\rho'$ are associated with corresponding $BE_k$-descriptors up to depth $n-1$, and vice versa;
  - given two tracks $\tilde{\rho}$ and $\tilde{\rho}'$ associated with $\mathcal{D}_{BE_k}$ and $\mathcal{D}'_{BE_k}$, respectively, for any track $\rho$, with $(v_{fin}, \text{fst}(\rho)) \in \delta$, there is a track $\rho'$, with $(v'_{fin}, \text{fst}(\rho')) \in \delta$, such that both $\tilde{\rho} \cdot \rho$ and $\tilde{\rho}' \cdot \rho'$ belong to $\text{Trk}_\mathcal{K}$, and they are associated with corresponding $BE_k$-descriptors up to depth $n-1$, and vice versa;
  - given two tracks $\tilde{\rho}$ and $\tilde{\rho}'$ associated with $\mathcal{D}_{BE_k}$ and $\mathcal{D}'_{BE_k}$, respectively, for any track $\rho$, with $(\text{lst}(\rho), v_{in}) \in \delta$, there is a track $\rho'$, with $(\text{lst}(\rho'), v'_{in}) \in \delta$, such that both $\rho \cdot \tilde{\rho}$ and $\rho' \cdot \tilde{\rho}'$ belong to $\text{Trk}_\mathcal{K}$, and they are associated with corresponding $BE_k$-descriptors up to depth $n-1$, and vice versa;
  - whenever $k > 0$, for any subtree of depth $k-1$ in $\mathcal{D}_{BE_k}$, whose root is linked to the root of $\mathcal{D}_{BE_k}$ via a *B*-edge (resp. *E*-edge), there is a subtree of depth $k-1$ in $\mathcal{D}'_{BE_k}$, whose root is linked to the root of $\mathcal{D}'_{BE_k}$ via a *B*-edge (resp., *E*-edge), corresponding up to depth $n-1$, and vice versa.

It can be easily checked that the correspondence between descriptors is an equivalence relation (reflexivity and symmetry are straightforward, while transitivity can be proved by induction on $n \geq 0$). Definition 17 expresses a form of (bounded) bisimulation among (the nodes associated with) the $BE_k$-descriptors in the graph $\mathscr{G}$ with respect to the defined relations of $A$-successor, $\overline{A}$-successor, $\overline{B}$-successor, $\overline{E}$-successor, and $B$- and $E$-subtrees.

For technical reasons, we need to introduce a variant of a previously-defined concept, the nesting depth of formulas, to take into consideration the nesting of all HS modalities (not only $B$ and $E$ as in Definition 7).

**Definition 18** The nesting depth of an HS formula $\psi$, denoted by $\text{Nest}(\psi)$, is inductively defined on the structure of $\psi$ as follows:

- $\text{Nest}(p) = 0$, for any proposition letter $p \in \mathcal{AP}$;
- $\text{Nest}(\neg \psi) = \text{Nest}(\psi)$;
- $\text{Nest}(\psi \wedge \varphi) = \max\{\text{Nest}(\psi), \text{Nest}(\varphi)\}$;
- $\text{Nest}(\langle X \rangle \psi) = 1 + \text{Nest}(\psi)$, with $X \in \{A, B, E, \overline{A}, \overline{B}, \overline{E}\}$.

It trivially holds that $\text{Nest}_{BE}(\psi) \leq \text{Nest}(\psi)$ for all HS formulas $\psi$.

We are now ready to state a couple of auxiliary lemmas preparatory to Theorem 3, whose proofs are given in the Appendix.

**Lemma 1** *Let $\mathcal{K} = (\mathcal{AP}, W, \delta, \mu, w_0)$ be a finite Kripke structure and $\rho, \rho'$ be two tracks in $\text{Trk}_{\mathcal{K}}$. For all $n, k \in \mathbb{N}$, with $k \leq n$, if $\mathcal{K}, \rho \models \varphi \iff \mathcal{K}, \rho' \models \varphi$ for all HS formulas $\varphi$ with $\text{Nest}_{BE}(\varphi) \leq k$ and $\text{Nest}(\varphi) \leq n$, then the $BE_k$-descriptors of $\rho$ and $\rho'$ are corresponding up to depth $n$.*

**Lemma 2** *Let $\mathcal{K} = (\mathcal{AP}, W, \delta, \mu, w_0)$ be a finite Kripke structure. For all $n, k \in \mathbb{N}$, with $k \geq 1$, if two descriptors $\mathcal{D}_{BE_k}$ and $\mathcal{D}'_{BE_k}$ are corresponding up to depth $n$, then $\mathcal{D}_{BE_k}|_{k-1}$ and $\mathcal{D}'_{BE_k}|_{k-1}$ are corresponding up to depth $n$ as well.*

As the last preparatory step, we provide a general definition of corresponding descriptors, where we remove the dependency on a specific depth $n$.

**Definition 19** Let $\mathcal{K}$ be a finite Kripke structure and let $\mathcal{D}_{BE_k}$ and $\mathcal{D}'_{BE_k}$ be two $BE_k$-descriptors associated with some of its tracks. We say that $\mathcal{D}_{BE_k}$ and $\mathcal{D}'_{BE_k}$ are *corresponding $BE_k$-descriptors* if and only if they are corresponding up to depth $n$, for all $n \in \mathbb{N}$.

It can be easily seen that it is an equivalence relation. Moreover, it is possible to show that two descriptors are corresponding if and only if the associated nodes in the graph $\mathscr{G}$ are bisimilar. Thus, the definition of correspondence between descriptors could be equivalently expressed in terms of a standard notion of bisimilarity among nodes of $\mathscr{G}$.

**Theorem 3** *Let $\mathcal{K}$ be a finite Kripke structure, $k \in \mathbb{N}$, and $\rho, \rho' \in \text{Trk}_{\mathcal{K}}$. The tracks $\rho$ and $\rho'$ are $k$-equivalent if and only if $\rho$ and $\rho'$ are associated with corresponding $BE_k$-descriptors.*

*Proof* ($\Rightarrow$) Let us first show that if $\rho$ and $\rho'$ are $k$-equivalent, then they are associated with corresponding $BE_k$-descriptors. The proof directly follows from Lemma 1. Since $\rho$ and $\rho'$ are $k$-equivalent, that is, $\mathcal{K}, \rho \models \psi \iff \mathcal{K}, \rho' \models \psi$ for all HS formulas $\psi$ with $\text{Nest}_{BE}(\psi) \leq k$ and no bound on $\text{Nest}(\psi)$, then their $BE_k$-descriptors are corresponding, with no bound on the depth of such a correspondence.

($\Leftarrow$) We now prove that, for any HS formula $\psi$, with $\text{Nest}_{BE}(\psi) = k$, if $\rho$ and $\rho'$ are associated with corresponding $BE_k$-descriptors, then they are $k$-equivalent, that is, $\mathcal{K}, \rho \models \psi \iff \mathcal{K}, \rho' \models \psi$. The proof is by induction on the structure of the formula.

- Let $\mathcal{K}, \rho \models p$, for some $p \in \mathcal{AP}$. Since the roots for the $BE$-descriptors of $\rho$ and $\rho'$ are labelled with the same set of proposition letters, it immediately follows that $\mathcal{K}, \rho' \models p$.
- Let $\mathcal{K}, \rho \models \psi_1 \wedge \psi_2$. Then, $\mathcal{K}, \rho \models \psi_1$ and $\mathcal{K}, \rho \models \psi_2$. Let $\text{Nest}_{BE}(\psi_1) = k$ and assume w.l.o.g. that $\text{Nest}_{BE}(\psi_2) = h \leq k$. By Definition 17 and Lemma 2, it immediately follows that if $\rho$ and $\rho'$ have corresponding $BE_k$-descriptors, then they also have corresponding $BE_h$-descriptors, with $h \leq k$. Hence, by the inductive hypothesis, $\mathcal{K}, \rho' \models \psi_1$ and $\mathcal{K}, \rho' \models \psi_2$, and, as a consequence, $\mathcal{K}, \rho' \models \psi_1 \wedge \psi_2$.
- Let $\mathcal{K}, \rho \models \neg \psi$. Then, $\mathcal{K}, \rho \not\models \psi$. By the inductive hypothesis, $\mathcal{K}, \rho' \not\models \psi$, and thus $\mathcal{K}, \rho' \models \neg \psi$.
- Let $\mathcal{K}, \rho \models \langle A \rangle \psi$. Then, there exists a track $\overline{\rho} \in \text{Trk}_\mathcal{K}$, with $\text{fst}(\overline{\rho}) = \text{lst}(\rho)$, such that $\mathcal{K}, \overline{\rho} \models \psi$. Since the $BE_k$-descriptors for $\rho$ and $\rho'$ are corresponding, there exists, in particular, a track $\overline{\rho}' \in \text{Trk}_\mathcal{K}$, with $\text{fst}(\overline{\rho}') = \text{lst}(\rho')$, such that the $BE_k$-descriptors for $\overline{\rho}$ and $\overline{\rho}'$ are corresponding. By the inductive hypothesis, $\mathcal{K}, \overline{\rho}' \models \psi$, so $\mathcal{K}, \rho' \models \langle A \rangle \psi$. The $\langle \overline{A} \rangle$ case is symmetric (notice that, due to the correspondence of the $BE_k$-descriptors for $\rho$ and $\rho'$, there exists $\overline{\rho} \in \text{Trk}_\mathcal{K}$, with $\text{lst}(\overline{\rho}) = \text{fst}(\rho)$, if and only if there exists $\overline{\rho}' \in \text{Trk}_\mathcal{K}$, with $\text{lst}(\overline{\rho}') = \text{fst}(\rho')$).
- Let $\mathcal{K}, \rho \models \langle \overline{B} \rangle \psi$. Then, there exists a track $\overline{\rho}$, with $(\text{lst}(\rho), \text{fst}(\overline{\rho})) \in \delta$ and $\rho \cdot \overline{\rho} \in \text{Trk}_\mathcal{K}$, such that $\mathcal{K}, \rho \cdot \overline{\rho} \models \psi$. Since the $BE_k$-descriptors for $\rho$ and $\rho'$ are corresponding, there exists, in particular, a track $\overline{\rho}'$, with $(\text{lst}(\rho'), \text{fst}(\overline{\rho}')) \in \delta$, such that $\rho \cdot \overline{\rho}$ and $\rho' \cdot \overline{\rho}' \in \text{Trk}_\mathcal{K}$ have corresponding $BE_k$-descriptors. By the inductive hypothesis $\mathcal{K}, \rho' \cdot \overline{\rho}' \models \psi$, and thus $\mathcal{K}, \rho' \models \langle \overline{B} \rangle \psi$. The $\langle \overline{E} \rangle$ case is symmetric (a remark similar to the one for the $\langle \overline{A} \rangle$ case can be done).
- Let $\mathcal{K}, \rho \models \langle B \rangle \psi$. Then, there exists a track $\overline{\rho} \in \text{Pref}(\rho)$ such that $\mathcal{K}, \overline{\rho} \models \psi$. Since the $BE_k$-descriptors for $\rho$ and $\rho'$ are corresponding, the subtree of depth $k-1$ for $\overline{\rho}$, in the $BE_k$-descriptor for $\rho$, corresponds to a subtree of depth $k-1$, in the $BE_k$-descriptor for $\rho'$. By definition of descriptor, there exists a track $\overline{\rho}' \in \text{Pref}(\rho')$ associated with the latter subtree. By the inductive hypothesis, $\mathcal{K}, \overline{\rho}' \models \psi$, and thus $\mathcal{K}, \rho' \models \langle B \rangle \psi$. The $\langle E \rangle$ case is symmetric, and thus its analysis is omitted.

This concludes the proof. □

We started the section by illustrating the case of the two tracks $v_0^5$ and $v_0^6$ of $\mathcal{K}_{Equiv}$. They have the same $BE_2$-descriptor (it is shown in Figure 8(a)), but not the same $BE_3$-descriptor (the $BE_3$-descriptor for $v_0^5$ is shown in Figure 8(b)). The $BE_3$-descriptor for $v_0^6$, indeed, features one more subtree, that is, the $BE_2$-descriptor for $v_0^5$, which is not present in Figure 8(b). However, such a subtree corresponds to the $BE_2$-descriptor for $v_0^4$. Symmetrically, the same happens for the suffix $v_0^5$ of $v_0^6$. Thus, $v_0^5$ and $v_0^6$, which have corresponding $BE_3$-descriptors, are 3-equivalent by Theorem 3. On the other hand, the $BE_4$-descriptors for $v_0^5$ and $v_0^6$ are not corresponding. In Figure 9, a part of the graph $\mathscr{G}$ of the *BE*-descriptors for the tracks of $\mathcal{K}_{Equiv}$ is shown. As it is evident from the figure, there exists a path consisting of 4 *B*-edges starting from the node of the $BE_4$-descriptor for $v_0^6$, whereas there is no a path of the same length starting from the node of the $BE_4$-descriptor for $v_0^5$. Hence, $v_0^5$ and $v_0^6$ are not 4-equivalent (as we already pointed out, $\mathcal{K}, v_0^6 \models \langle B \rangle^4 \top$, while $\mathcal{K}, v_0^5 \not\models \langle B \rangle^4 \top$).

## 5 EXPSPACE-hardness

We conclude the paper by proving that the model checking problem for HS, against finite Kripke structures, is EXPSPACE-hard. As a preparatory work, we introduce a succinct encoding of HS formulas, according to which we write $\langle B \rangle^k \psi$ for

$$\underbrace{\langle B \rangle \langle B \rangle \cdots \langle B \rangle}_{k \text{ times}} \psi,$$

and we represent $k$ in binary (the same for all the other HS modalities). As we will prove, if we exploit this encoding, the model checking problem for HS is EXPSPACE-hard, otherwise—using the standard unary notation—it is PSPACE-hard.

**Theorem 4** *The model checking problem for HS against finite Kripke structures is EXPSPACE-hard (under a LOGSPACE reduction), if formulas are* succinctly en-coded, *otherwise it is PSPACE-hard.*

*Proof* Let us consider a language $L$ decided by a *deterministic one-tape* Turing machine $M$ (w.l.o.g.) that, on an input of size $n$, requires no more than $2^{n^k} - 3$ symbols on its tape (we are assuming a high enough constant $k \in \mathbb{N}$). Hence, $L$ belongs to EXPSPACE. Let $\Sigma$ and $Q$ be respectively the alphabet and the set of states of $M$, and let # be a special symbol, which does not belong to $\Sigma$, used as separator for configurations (in the following, we let $\Sigma' = \Sigma \cup \{\#\}$). The alphabet $\Sigma$ is assumed to contain the blank symbol $\sqcup$. As usual, a computation of $M$ is a sequence of configurations of $M$, where each configuration fixes the content of the tape, the position of the head on the tape, and the internal state of $M$.

We exploit a standard encoding for computations, called *computation table* (or tableau) (see [21, 26] for further details). Each configuration of $M$ is a sequence over the alphabet $\Gamma = \Sigma' \cup (Q \times \Sigma)$. A symbol $(q, c) \in Q \times \Sigma$ occurring at the $i$-th position encodes the fact that the machine has an internal state $q$ and its head is currently on the $i$-th position of the tape (obviously, there is exactly one occurrence of a symbol in $Q \times \Sigma$ in each configuration).

Since $M$ uses no more than $2^{n^k} - 3$ symbols on its tape, the size of a configuration is $2^{n^k}$ (we need 3 occurrences of the special symbol #, two for delimiting the beginning of the configuration and one for the end; additionally, $M$ never overwrites delimiters #). If a configuration is actually shorter than $2^{n^k}$, it is padded with $\sqcup$ symbols to reach length $2^{n^k}$ (which is a fixed number, once the input length is known).

The computation table is a matrix of $2^{n^k}$ columns, where the $i$-th row records the configuration of $M$ at the $i$-th computation step.

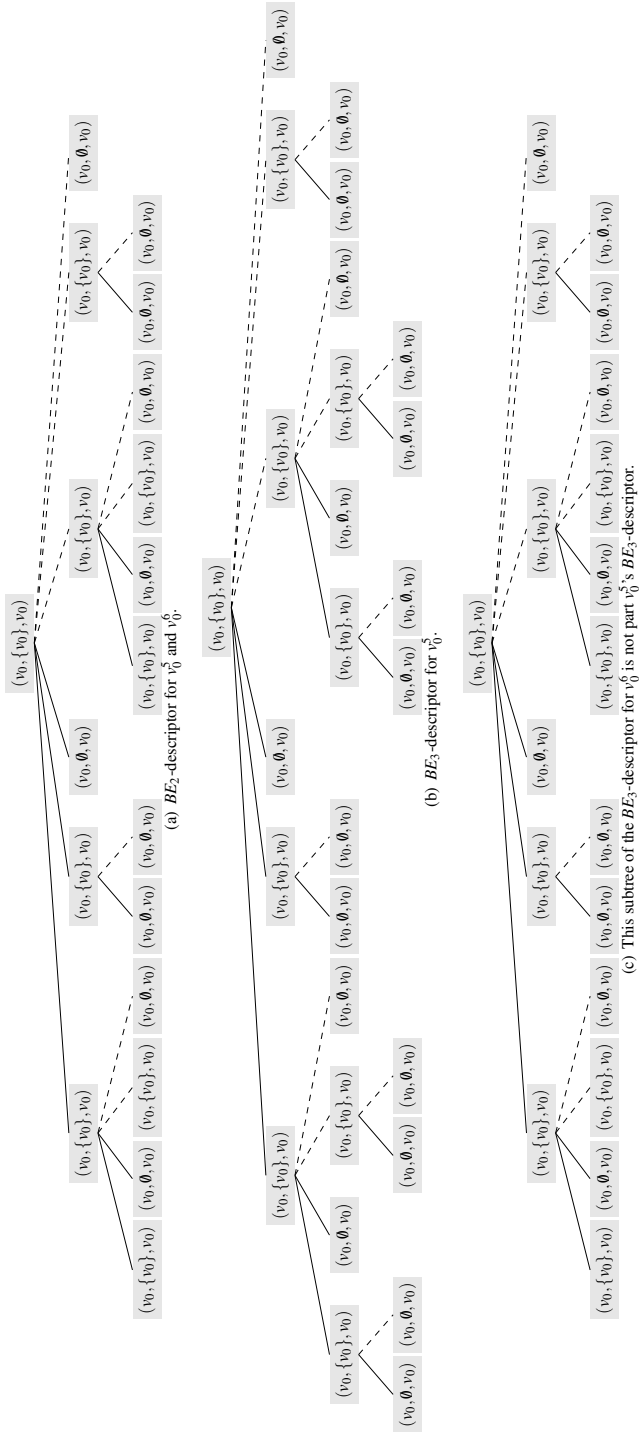

**Fig. 8** Tracks $v_0^5$ and $v_0^6$ have the same $BE_2$-descriptor, which is shown in Figure 8(a). Figure 8(b) depicts the $BE_3$-descriptor for $v_0^5$. The $BE_3$-descriptor for $v_0^6$ is not shown due to lack of space. With respect to the one in Figure 8(b), it features two additional subtrees, one associated with a prefix, the other with a suffix. The former subtree is shown in 8(c); not surprisingly, it is isomorphic (equal) to 8(a).

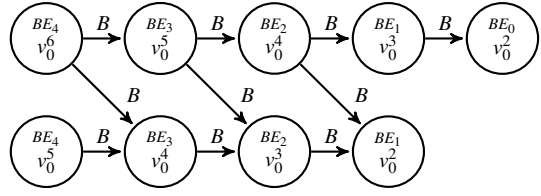

**Fig. 9** Part of the graph $\mathscr{G}$ of the $BE_t$-descriptors ($t \leq 4$) for the tracks of $\mathcal{K}_{Equiv}$. In each node, we report the depth of the descriptors they are associated with (top) and a witness track for the descriptor (bottom).

| # | # | $(q_0,c_0)$ | $c_1$ | $c_2$ | $\cdots$ | $\cdots$ | $c_{n-1}$ | ⊔ | ⊔ | $\cdots$ | $\cdots$ | ⊔ | # |
|---|---|---|---|---|---|---|---|---|---|---|---|---|---|
| # | # | $c'_0$ | $(q_1,c_1)$ | $c_2$ | $\cdots$ | $\cdots$ | $c_{n-1}$ | ⊔ | ⊔ | $\cdots$ | $\cdots$ | ⊔ | # |
| $\vdots$ | $\vdots$ | | | | $\ddots$ | $\ddots$ | | | | | | | $\vdots$ |
| $\vdots$ | $\vdots$ | | | | $\ddots$ | $\ddots$ | | | | | | | $\vdots$ |
| # | # | $\cdots$ | $\cdots$ | $(q_{yes},c_k)$ | $\cdots$ | $\cdots$ | $\cdots$ | $\cdots$ | $\cdots$ | $\cdots$ | $\cdots$ | $\cdots$ | # |

$$\underbrace{\phantom{}}_{2^{n^k}}$$

**Fig. 10** An example of a computation table.

An example of a table is given in Figure 10. In the first configuration (row), the head is in the leftmost position (to the right of the delimiters #) and $M$ is in state $q_0$. In addition, the string symbols $c_0 c_1 \cdots c_{n-1}$ are padded with occurrences of ⊔'s to reach length $2^{n^k}$. In the second configuration, the head has moved one position to the right, $c_0$ has been overwritten with $c'_0$, and $M$ is in state $q_1$. From the first two rows, we can deduce that the tuple $(q_0, c_0, q_1, c'_0, \rightarrow)$ belongs to the transition relation $\delta_M \subseteq Q \times \Sigma \times Q \times \Sigma \times \{\rightarrow, \leftarrow, \bullet\}$ of $M$, with the standard meaning for the components (the first one gives the current state, the second the symbol on tape currently read, the third the next state, the fourth the symbol replaced in the current position, the fifth the move of the head to right, left, or stay). Being $M$ deterministic, $\delta_M$ is actually a function of $Q \times \Sigma$.

Following [21,26], we introduce the notion of (legal) window. A window is a $2 \times 3$ matrix, in which the first row represents three consecutive symbols of a possible configuration. The second row represents the three symbols which are placed exactly in the same position in the next configuration. A window is legal when the changes from the first to the second row are coherent with $\delta_M$ in the obvious sense. Actually, the set of legal windows, which we denote by $Wnd \subseteq \left(\Gamma^3\right)^2$, is a suitable tabular representation of the transition relation $\delta_M$. For instance, two legal windows associated with the table of the previous example are:

| # | $(q_0,c_0)$ | $c_1$ |
|---|---|---|
| # | $c'_0$ | $(q_1,c_1)$ |

| $(q_0,c_0)$ | $c_1$ | $c_2$ |
|---|---|---|
| $c'_0$ | $(q_1,c_1)$ | $c_2$ |

Formally, a pair $((x,y,z),(x',y',z')) \in Wnd$ can be represented as follows:

| $x$ | $y$ | $z$ |
|---|---|---|
| $x'$ | $y'$ | $z'$ |

    with $x,x',y,y',z,z' \in \Gamma$,

where the following constraints must be satisfied:

1. if all $x, y, z \in \Sigma'$ ($x, y, z$ are not state-symbol pairs), then $y = y'$;
2. if one among $x$, $y$, and $z$ belongs to $Q \times \Sigma$, then $x'$, $y'$ and $z'$ are univocally determined by $\delta_M$;
3. $(x = \# \Rightarrow x' = \#) \wedge (y = \# \Rightarrow y' = \#) \wedge (z = \# \Rightarrow z' = \#)$.

As we already said, $M$ never overwrites an occurrence of #; we can assume that the head never visits a cell labelled with # as well (see [21]). As a matter of fact, in some window, condition 2 would require to move the head right (or left) overwriting # (or just visiting it), while 3 does not allow one to replace an occurrence of # with another symbol (notice that $(q_i, \#)$ does not belong to $\Gamma$ for any state $q_i$ of $M$). In such a case, the window is not valid and thus it is discarded (it does not belong to $Wnd$).

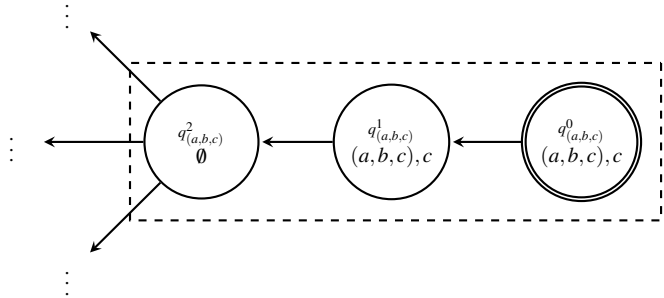

**Fig. 11** An instance of the gadget for $(a, b, c) \in \Gamma^3$.

In the following, we define a finite Kripke structure $\mathcal{K} = (\mathcal{AP}, W, \delta, \mu, w_0)$ and an $HS[A, \overline{A}, B, \overline{B}, E, \overline{E}]$ formula $\psi$ such that $\mathcal{K} \models \psi$ if and only if $M$ accepts its input string $c_0 c_1 \cdots c_{n-1}$. The set of proposition letters is $\mathcal{AP} = \Gamma \cup \Gamma^3 \cup \{start\}$. The finite Kripke structure $\mathcal{K}$ is obtained by suitably composing a basic pattern, called *gadget* (see Figure 11). Any instance of the gadget is associated with a triple of symbols $(a, b, c) \in \Gamma^3$, that is, a sequence of three adjacent symbols in a configuration, and it consists of 3 states $q^0_{(a,b,c)}$, $q^1_{(a,b,c)}$, and $q^2_{(a,b,c)}$ such that

$$\mu(q^0_{(a,b,c)}) = \mu(q^1_{(a,b,c)}) = \{(a,b,c), c\} \text{ and } \mu(q^2_{(a,b,c)}) = \emptyset.$$

Moreover,
$$\delta(q^0_{(a,b,c)}) = \{q^1_{(a,b,c)}\} \text{ and } \delta(q^1_{(a,b,c)}) = \{q^2_{(a,b,c)}\}.$$

The underlying idea is that a gadget associated with $(x, y, z) \in \Gamma^3$ "records" the current proposition letter $z$ and the two "past" (immediately preceding) proposition letters $x$ and $y$.

The finite Kripke structure $\mathcal{K}$ has (an instance of) a gadget for every $(x, y, z) \in \Gamma^3$, and for all $(x, y, z), (x', y', z') \in \Gamma^3$, it holds that $q^0_{(x',y',z')} \in \delta(q^2_{(x,y,z)})$ if and only if $x' = y$ and $y' = z$. Moreover, $\mathcal{K}$ has some additional (auxiliary) states $w_0, \cdots, w_6$, whose relationships are described in Figure 12, and $\delta(w_6) = \{q^0_{(\#,\#,x)} \mid x \in \Gamma\}$. It is worth noticing that the overall size of $\mathcal{K}$ only depends on $|\Gamma|$ and it is constant with respect to the input string $c_0 c_1 \cdots c_{n-1}$ of $M$.

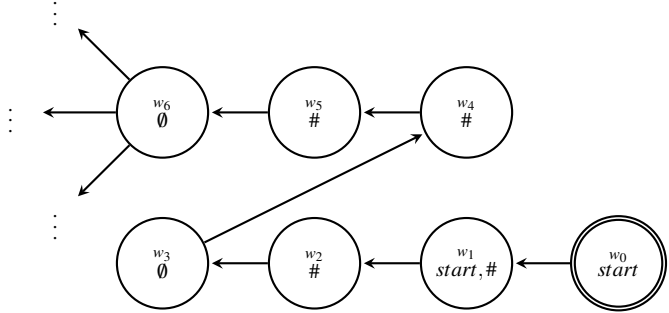

**Fig. 12** Initial part of $\mathcal{K}$.

Now, we want to decide whether or not an input string belongs to the language $L$ by solving the model checking problem $\mathcal{K} \models \textit{start} \to \langle A \rangle \xi$, where $\xi$ is satisfied only by those tracks which represent a successful computation of $M$. Since the only (initial) track which satisfies *start* is $w_0 w_1$ (see Figure 12), we are actually verifying the existence of a track which begins with $w_1$ and satisfies $\xi$.

As for $\xi$, it basically requires that a track $\rho$, with $\text{fst}(\rho) = w_1$, for which $\mathcal{K}, \rho \models \xi$, mimics a successful computation of $M$. First, every interval $\rho(i, i+1)$, with $i \mod 3 = 0$, satisfies the proposition letter $p \in \mathcal{AP}$ if and only if the $\frac{i}{3}$-th character of the computation represented by $\rho$ is $p$ (notice that as a consequence of the gadget structure, only subtracks $\overline{\rho} = \rho(i, i+1)$, with $i \mod 3 = 0$, of $\rho$ can satisfy some proposition letters). A symbol of a configuration is mapped to an occurrence of an instance of a gadget in $\rho$; in turn, $\rho$ encodes a computation of $M$ through the concatenation of the first, second, third... rows of the computation table (two consecutive configurations are separated by 3 occurrences of #, which require 9 states overall).

Let us now formally define the HS formula $\xi$:

$$\xi = \psi_{\textit{accept}} \wedge \psi_{\textit{input}} \wedge \psi_{\textit{window}}.$$

The first conjunct

$$\psi_{\textit{accept}} = \langle B \rangle \langle A \rangle \bigvee_{a \in \Sigma} (q_{\textit{yes}}, a)$$

requires a track to contain an occurrence of the accepting state of $M$ $q_{\textit{yes}}$.

The second conjunct $\psi_{\textit{input}}$ is a bit more involved. It requires that the subtrack corresponding to the first configuration of $M$ actually "spells" the input $c_0 c_1 \cdots c_{n-1}$, suitably padded with occurrences of $\sqcup$ and ended by a # (we recall that the formula $\ell(k)$, which has been introduced in Section 2, is satisfied only by those tracks whose length equals $k$, with $k \geq 2$, and it has a binary encoding of $O(\log k)$ bits):

$$\psi_{input} = [B](\ell(7) \to \langle A \rangle (q_0, c_0)) \wedge [B](\ell(10) \to \langle A \rangle c_1) \wedge [B](\ell(13) \to \langle A \rangle c_2) \wedge$$
$$\vdots$$
$$[B](\ell(7+3(n-1)) \to \langle A \rangle c_{n-1}) \wedge$$
$$[B](\langle B \rangle^{5+3n} \top \wedge [B]^{3 \cdot 2^{n^k}-6} \bot \to \langle A \rangle ((\ell(2) \wedge \bigwedge_{a \in \Gamma} \neg a) \vee \sqcup)) \wedge$$
$$[B](\ell(3 \cdot 2^{n^k} - 2) \to \langle A \rangle \#).$$

Finally, the third conjunct $\psi_{window}$ enforces the window constraint: if the proposition $(d,e,f) \in \Gamma^3$ is witnessed by a subinterval (of length 2) in the subtrack of $\rho$ corresponding to the $j$-th configuration of $M$, then, at the same position of (the subtrack of $\rho$ associated with) configuration $j-1$, there must be some $(a,b,c) \in \Gamma^3$ such that $((a,b,c),(d,e,f)) \in Wnd$.

$$\psi_{window} = [B](\langle B \rangle^{3(2^{n^k}+2)+1} \top \to$$
$$\bigwedge_{(d,e,f) \in \Gamma^3} (\langle A \rangle (d,e,f) \to [E](\ell(3 \cdot 2^{n^k}) \to \bigvee_{((a,b,c),(d,e,f)) \in Wnd} \langle \overline{A} \rangle (a,b,c)))).$$

The subformula $\langle B \rangle^{3(2^{n^k}+2)+1} \top$ guarantees that we are not considering the (subtrack associated with the) first configuration. Moreover, if some prefix $\tilde{\rho}$ of $\rho$ satisfies $(\langle B \rangle^{3(2^{n^k}+2)+1} \top$ and$)$ $\langle A \rangle (d,e,f)$, for some $(d,e,f) \in \Gamma^3$, then it holds that $\mathcal{K}, \tilde{\rho} \models [E](\ell(3 \cdot 2^{n^k}) \to \bigvee_{((a,b,c),(d,e,f)) \in Wnd} \langle \overline{A} \rangle (a,b,c)))$. This amounts to say that the suffix $\hat{\rho}$ of $\tilde{\rho}$ of length $3 \cdot 2^{n^k}$ is such that $\mathcal{K}, \hat{\rho} \models \bigvee_{((a,b,c),(d,e,f)) \in Wnd} \langle \overline{A} \rangle (a,b,c)$, that is, $\hat{\rho}$ is the subtrack between (the prefixes of $\rho$ corresponding to) the same position (same column) in two adjacent configurations (rows of the table), and it is forced to begin with an occurrence of $q^1_{(a,b,c)}$ and to end with $q^0_{(d,e,f)}$, for some $((a,b,c),(d,e,f)) \in Wnd$.

It is immediate to check that all the integers which need to be stored in the formula are less than or equal to $3 \cdot 2^{n^k} + 7$, and thus $O(n^k)$ bits suffice. This allows us to conclude that the formula can be generated in polynomial time (and logarithmic working space).

If we *do not* allow the binary encoding of the exponents, the model checking problem for HS formulas is PSPACE-hard (under a LOGSPACE reduction): the proof is the same as before, but in order for the formula $\xi$ to be generated in polynomial time, we must restrict ourselves to computations of Turing machines using at most polynomial space. $\square$

## 6 Related work

While the satisfiability problem for interval temporal logics has been extensively and systematically investigated in the literature [2, 8, 11, 12, 14, 20, 29], a little work has been done on model checking.

In [18], Montanari et al. give a first characterization of the model checking problem for full HS, interpreted over finite Kripke structures (under the homogeneity assumption). In that paper, the authors provide the basic elements of the general picture, namely, the interpretation of HS formulas over (abstract) interval models, the mapping of finite Kripke structures into (abstract) interval models, the notion of track descriptor, and a small model theorem proving the decidability of the model checking problem for full HS against finite Kripke structures. However, due to space constraints, technical details of the proofs are not fully worked out. Moreover, they do not provide any lower bound to the complexity of the problem (no hardness result is given), and the outlined model checking procedure for the fragments $HS[A,\overline{A},B,\overline{B}]$ and $HS[A,\overline{A},E,\overline{E}]$, based on the notion of compact track descriptor, is flawed.

In [15, 16], Lomuscio and Michaliszyn address the model checking problem for some fragments of HS extended with epistemic modalities. Their semantic assumptions differ from those made in [18], making it difficult to compare the outcomes of the two research directions. In both cases, formulas of interval temporal logic are evaluated over finite paths/tracks obtained from the unravelling of a finite Kripke structure. However, in [18] the authors state that a proposition letter holds over an interval (track) if and only if it holds over all its states (homogeneity principle), while in [15, 16] truth of proposition letters is defined over pairs of states (the endpoints of tracks/intervals).

In [15], the authors focus their attention on the HS fragment $HS[B,E,D]$ (since modality $\langle D \rangle$ is easily definable in terms of modalities $\langle B \rangle$ and $\langle E \rangle$, $HS[B,E,D]$ is actually as expressive as $HS[B,E]$), extended with epistemic modalities. They consider a restricted form of model checking, which verifies the given specification against a single (finite) initial computation interval. Their goal is indeed to reason about a given computation of a multi-agent system, rather than on all its admissible computations. The authors prove that the considered model checking problem is PSPACE-complete. Moreover, they show that the same problem restricted to the purely temporal fragment $HS[B,E,D]$, that is, the one obtained by removing epistemic modalities, is in PTIME. These results do not come as a surprise as they trade expressiveness for efficiency: modalities $B$ and $E$ allow one to access only sub-intervals of the initial one, whose number is quadratic in the length (number of states) of the initial interval.

In [16], they show that the picture drastically changes with other fragments of HS, that allow one to access infinitely many tracks/intervals. In particular, they prove that the model checking problem for the HS fragment $HS[A,\overline{B},L]$ (since modality $\langle L \rangle$ is easily definable in terms of modality $\langle A \rangle$, $HS[A,\overline{B},L]$ is actually as expressive as $HS[A,\overline{B}]$), extended with epistemic modalities, is decidable, with a non-elementary upper bound. Notice that, thanks to modalities $\langle A \rangle$ and $\langle \overline{B} \rangle$, formulas of this logic can possibly refer to infinitely many (future) tracks/intervals.

# 7 Conclusions and future work

In this paper, we devised a non-elementary model checking algorithm for full HS. Its cornerstone is the notion of $BE_k$-descriptor, which allows us to obtain a finite representation of a possibly infinite set of equivalent tracks. Since the number of $BE_k$-descriptors is always finite, the decidability of the model checking problem for HS over finite Kripke structures easily follows. In addition, we proved that such a problem is EXPSPACE-hard, provided that a succinct encoding of formulas is used (otherwise, we can only prove that it is PSPACE-hard).

We are exploring the possibility of obtaining (much) more efficient model checking algorithms by restricting to suitable fragments of HS. In particular, we are studying the effects of the removal of the modality $E$ (resp., $B$) from HS. More precisely, we are thinking of the possibility of applying to $HS[A,\overline{A},B,\overline{B},\overline{E}]$ and $HS[A,\overline{A},E,\overline{E},\overline{B}]$ a contraction method to restrict the verification of the formula to a finite subset of tracks of bounded size. Other HS fragments of interest are $HS[A,\overline{A},\overline{B},\overline{E}]$, that we conjecture to be PSPACE-complete, and the "orthogonal" fragment $HS[A,\overline{A},B,E]$. Another interesting fragment is $HS[A,\overline{A}]$ (the logic of temporal neighbourhood): it can easily be shown that its model checking problem is NP-hard, but we can only think of PSPACE algorithms.

Last but not least, it is worth exploring the model checking problem for HS and its fragments under other semantic interpretations (relaxing the homogeneity assumption). Moreover, we are thinking of the possibility of replacing finite Kripke structures by richer computational models such as game-theoretic and/or infinite state structures. These models have been extensively exploited in formal verification with classical temporal logics, and we expect them to be quite beneficial in the interval setting.

**Acknowledgements**

We would like to thank the anonymous reviewers whose comments and suggestions helped us to improve the paper. Angelo Montanari, Aniello Murano, and Adriano Peron acknowledge the support from the GNCS project: "Algoritmica for model checking and synthesis of safety-critical systems". Aniello Murano and Adriano Peron also acknowledge the support from the FP7 EU project 600958-SHERPA.

**Appendix**

Proof of Lemma 1.

*Proof* The proof is by induction on $n \geq 0$. Let $\mathcal{D}_{BE_k}$ and $\mathcal{D}'_{BE_k}$ be the $BE_k$-descriptors for $\rho$ and $\rho'$, respectively.

Base case ($n = 0$). Since $\mathcal{K}, \rho \models p \iff \mathcal{K}, \rho' \models p$, for any $p \in \mathcal{AP}$, the roots of $\mathcal{D}_{BE_k}$ and $\mathcal{D}'_{BE_k}$ are labelled by the same set of proposition letters and the descriptors are corresponding up to depth 0.

Inductive step ($n \geq 1$). We preliminarily show that if $\mathcal{K}, \rho \models \varphi \iff \mathcal{K}, \rho' \models \varphi$ for all HS formulas $\varphi$ with $\mathrm{Nest}_{BE}(\varphi) \leq k$ and $\mathrm{Nest}(\varphi) \leq n$, then for any track $\overline{\rho} \in \mathrm{Trk}_{\mathcal{K}}$, with $\mathrm{fst}(\overline{\rho}) = \mathrm{lst}(\rho)$, there is a track $\overline{\rho}' \in \mathrm{Trk}_{\mathcal{K}}$, with $\mathrm{fst}(\overline{\rho}') = \mathrm{lst}(\rho')$, such that, for all HS formulas $\psi$, with $\mathrm{Nest}(\psi) \leq n - 1$ and $\mathrm{Nest}_{BE}(\psi) \leq k$, $\mathcal{K}, \overline{\rho} \models \psi \iff \mathcal{K}, \overline{\rho}' \models \psi$. The proof is by contradiction. Suppose that there exists a track $\overline{\rho} \in \mathrm{Trk}_{\mathcal{K}}$, with $\mathrm{fst}(\overline{\rho}) = \mathrm{lst}(\rho)$, such that, for all tracks $\overline{\rho}' \in \mathrm{Trk}_{\mathcal{K}}$, with $\mathrm{fst}(\overline{\rho}') = \mathrm{lst}(\rho')$, there exists a formula $\psi$, with $\mathrm{Nest}(\psi) \leq n - 1$ and $\mathrm{Nest}_{BE}(\psi) \leq k$, such that $\mathcal{K}, \overline{\rho} \models \psi$ and $\mathcal{K}, \overline{\rho}' \not\models \psi$. Let $H$ be the set of those tracks $\hat{\rho}$ such that $\mathrm{fst}(\hat{\rho}) = \mathrm{lst}(\rho')$. $H$ can be partitioned into a finite number of classes, say $s \geq 1$, each one containing $k$-descriptor equivalent tracks of $H$ (remind that $k$-descriptor equivalence is an equivalence relation of finite index). Now, let $\{\overline{\rho}'_1, \overline{\rho}'_2, \ldots, \overline{\rho}'_s\}$ be a set of track representatives, chosen one for each equivalence class induced by $\sim_k$ on $H$ (for all $1 \leq i < j \leq s$, $\overline{\rho}'_i$ and $\overline{\rho}'_j$ have distinct $BE_k$-descriptors). By Theorem 1, tracks which are $k$-descriptor equivalent satisfy the same set of formulas $\psi'$, with $\mathrm{Nest}_{BE}(\psi') \leq k$. So there are formulas $\psi_1, \ldots, \psi_s$ such that, for all $1 \leq i \leq s$, $\mathrm{Nest}(\psi_i) \leq n - 1$, $\mathrm{Nest}_{BE}(\psi_i) \leq k$, $\mathcal{K}, \overline{\rho} \models \psi_i$, and $\mathcal{K}, \overline{\rho}'_i \not\models \psi_i$. It easily follows that $\mathcal{K}, \overline{\rho} \models \psi_1 \wedge \psi_2 \wedge \cdots \wedge \psi_s$ and, for all $1 \leq i \leq s$, $\mathcal{K}, \overline{\rho}'_i \models \neg \psi_1 \vee \neg \psi_2 \vee \cdots \vee \neg \psi_s$. Hence, $\mathcal{K}, \rho \models \langle A \rangle (\psi_1 \wedge \psi_2 \wedge \cdots \wedge \psi_s)$ and $\mathcal{K}, \rho' \models [A](\neg \psi_1 \vee \neg \psi_2 \vee \cdots \vee \neg \psi_s)$, that is, $\mathcal{K}, \rho' \not\models \langle A \rangle (\psi_1 \wedge \psi_2 \wedge \cdots \wedge \psi_s)$, which is a contradiction.

Thus, we have proved that for any track $\overline{\rho} \in \mathrm{Trk}_{\mathcal{K}}$, with $\mathrm{fst}(\overline{\rho}) = \mathrm{lst}(\rho)$, there exists a track $\overline{\rho}' \in \mathrm{Trk}_{\mathcal{K}}$, with $\mathrm{fst}(\overline{\rho}') = \mathrm{lst}(\rho')$, such that, for all HS formulas $\psi$, with $\mathrm{Nest}(\psi) \leq n - 1$ and $\mathrm{Nest}_{BE}(\psi) \leq k$, $\mathcal{K}, \overline{\rho} \models \psi \iff \mathcal{K}, \overline{\rho}' \models \psi$. By the inductive hypothesis, $\overline{\rho}$ and $\overline{\rho}'$ are associated with corresponding $BE_k$-descriptors up to depth $n - 1$. Symmetrically, we can show that for any track $\overline{\rho}' \in \mathrm{Trk}_{\mathcal{K}}$, with $\mathrm{fst}(\overline{\rho}') = \mathrm{lst}(\rho')$, there exists $\overline{\rho} \in \mathrm{Trk}_{\mathcal{K}}$, with $\mathrm{fst}(\overline{\rho}) = \mathrm{lst}(\rho)$, such that $\overline{\rho}'$ and $\overline{\rho}$ are associated with corresponding $BE_k$-descriptors up to depth $n - 1$. In this way, we have proved the condition for modality $A$ of Definition of 17. The conditions for modalities $\overline{A}$, $\overline{B}$, and $\overline{E}$ can be proved in a very similar way. In particular, as a consequence of the fact that $\mathcal{K}, \rho \models \varphi \iff \mathcal{K}, \rho' \models \varphi$ for all HS formulas $\varphi$ with $\mathrm{Nest}_{BE}(\varphi) \leq k$ and $\mathrm{Nest}(\varphi) \leq n$, with $n \geq 1$, it holds that $\mathcal{K}, \rho \models \langle \overline{A} \rangle \top \iff \mathcal{K}, \rho' \models \langle \overline{A} \rangle \top$. It follows that $\mathcal{D}_{BE_k}$ has an $\overline{A}$-successor if and only if $\mathcal{D}'_{BE_k}$ has one. The same holds for $\overline{E}$-successors.

Let us now consider the condition for modality $B$ of Definition of 17.

First of all, we show that for any track $\overline{\rho} \in \mathrm{Pref}(\rho)$, there exists a track $\overline{\rho}' \in \mathrm{Pref}(\rho')$ such that for all HS formulas $\psi$, with $\mathrm{Nest}(\psi) \leq n - 1$ and $\mathrm{Nest}_{BE}(\psi) \leq k - 1$, $\mathcal{K}, \overline{\rho} \models \psi \iff \mathcal{K}, \overline{\rho}' \models \psi$. The proof is again by contradiction. Suppose that there exists a track $\overline{\rho} \in \mathrm{Pref}(\rho)$ such that, for all tracks $\overline{\rho}' \in \mathrm{Pref}(\rho')$, there exists a formula $\psi$, with $\mathrm{Nest}(\psi) \leq n - 1$ and $\mathrm{Nest}_{BE}(\psi) \leq k - 1$, such that $\mathcal{K}, \overline{\rho} \models \psi$ and $\mathcal{K}, \overline{\rho}' \not\models \psi$. Now, let us consider the tracks $\overline{\rho}'_1, \overline{\rho}'_2, \cdots, \overline{\rho}'_s$ (for some $s \in \mathbb{N}$) which are prefixes of $\rho'$ and are associated with distinct subtrees of depth $k - 1$ of the $BE_k$-descriptor for $\rho'$ (the number of these tracks is obviously finite). So there are formulas $\psi_1, \ldots, \psi_s$ such that, for all $1 \leq i \leq s$, $\mathrm{Nest}(\psi_i) \leq n - 1$, $\mathrm{Nest}_{BE}(\psi_i) \leq k - 1$, $\mathcal{K}, \overline{\rho} \models \psi_i$, and $\mathcal{K}, \overline{\rho}'_i \not\models \psi_i$. Thus, $\mathcal{K}, \overline{\rho} \models \psi_1 \wedge \psi_2 \wedge \cdots \wedge \psi_s$ and for all $i$, $\mathcal{K}, \overline{\rho}'_i \models \neg \psi_1 \vee \neg \psi_2 \vee \cdots \vee \neg \psi_s$.

Hence $\mathcal{K}, \rho \models \langle B \rangle (\psi_1 \wedge \psi_2 \wedge \cdots \wedge \psi_s)$ and $\mathcal{K}, \rho' \models [B](\neg \psi_1 \vee \neg \psi_2 \vee \cdots \vee \neg \psi_s)$, that is $\mathcal{K}, \rho' \not\models \langle B \rangle (\psi_1 \wedge \psi_2 \wedge \cdots \wedge \psi_s)$, which leads to a contradiction.

We have proved that for any track $\overline{\rho} \in \text{Pref}(\rho)$, there exists a track $\overline{\rho}' \in \text{Pref}(\rho')$ such that, for all HS formulas $\psi$, with $\text{Nest}(\psi) \leq n-1$ and $\text{Nest}_{BE}(\psi) \leq k-1$, $\mathcal{K}, \overline{\rho} \models \psi \iff \mathcal{K}, \overline{\rho}' \models \psi$. By the inductive hypothesis, $\overline{\rho}$ and $\overline{\rho}'$ are associated with corresponding $BE_{k-1}$-descriptors up to depth $n-1$. Symmetrically, we can show that for any track $\overline{\rho}' \in \text{Pref}(\rho')$, there exists a track $\overline{\rho} \in \text{Pref}(\rho)$ such that $\overline{\rho}'$ and $\overline{\rho}$ are associated with corresponding $BE_{k-1}$-descriptors up to depth $n-1$.

In this way, we have proved the condition for modality $B$ of Definition of 17. The condition for modality $E$ can be proved in a symmetrical way. □

Proof of Lemma 2.

*Proof* The proof is by induction on $n \geq 0$.

Base case ($n = 0$). Consider the descriptors $\mathcal{D}_{BE_k}$, $\mathcal{D}'_{BE_k}$, $\mathcal{D}_{BE_k}|_{k-1}$, and $\mathcal{D}'_{BE_k}|_{k-1}$. Since the roots of $\mathcal{D}_{BE_k}$ and $\mathcal{D}'_{BE_k}$ are labelled by the same set of proposition letters, the roots of $\mathcal{D}_{BE_k}|_{k-1}$ and $\mathcal{D}'_{BE_k}|_{k-1}$ are labelled by the same set of proposition letters as well.

Inductive step ($n > 0$). Let $\rho, \rho' \in \text{Trk}_\mathcal{K}$ be two witnesses for $\mathcal{D}_{BE_k}$ and for $\mathcal{D}'_{BE_k}$, respectively (and thus for $\mathcal{D}_{BE_k}|_{k-1}$ and and $\mathcal{D}'_{BE_k}|_{k-1}$, respectively). Consider a track $\tilde{\rho} \in \text{Trk}_\mathcal{K}$, with $\text{fst}(\tilde{\rho}) = \text{lst}(\rho)$. The $BE_k$-descriptor $\mathcal{D}_{BE_k}^{\tilde{}}$ for $\tilde{\rho}$ is an $A$-successor of $\mathcal{D}_{BE_k}$, and $\mathcal{D}_{BE_k}^{\tilde{}}|_{k-1}$ is an $A$-successor of $\mathcal{D}_{BE_k}|_{k-1}$. Since $\mathcal{D}_{BE_k}$ and $\mathcal{D}'_{BE_k}$ are corresponding up to depth $n$, there exists a track $\overline{\rho} \in \text{Trk}_\mathcal{K}$, with $\text{fst}(\overline{\rho}) = \text{lst}(\rho')$, described by $\overline{\mathcal{D}_{BE_k}}$, such that $\mathcal{D}_{BE_k}^{\tilde{}}$ and $\overline{\mathcal{D}_{BE_k}}$ are corresponding up to depth $n-1$. By the inductive hypothesis, $\mathcal{D}_{BE_k}^{\tilde{}}|_{k-1}$ and $\overline{\mathcal{D}_{BE_k}}|_{k-1}$ are corresponding up to depth $n-1$ (and, obviously, $\overline{\mathcal{D}_{BE_k}}|_{k-1}$ is an $A$-successor of $\mathcal{D}'_{BE_k}|_{k-1}$).

Let us consider now a track $\hat{\rho}$, with $(\text{lst}(\rho), \text{fst}(\hat{\rho})) \in \delta$ and $\rho \cdot \hat{\rho} \in \text{Trk}_\mathcal{K}$. The $BE_k$-descriptor $\mathcal{D}_{BE_k}^{\hat{}}$ of $\rho \cdot \hat{\rho}$ is a $\overline{B}$-successor of $\mathcal{D}_{BE_k}$ and $\mathcal{D}_{BE_k}^{\hat{}}|_{k-1}$ is a $\overline{B}$-successor of $\mathcal{D}_{BE_k}|_{k-1}$. Since $\mathcal{D}_{BE_k}$ and $\mathcal{D}'_{BE_k}$ are corresponding up to depth $n$, there exists a track $\check{\rho}$ such that $(\text{lst}(\rho'), \text{fst}(\check{\rho})) \in \delta$, $\rho' \cdot \check{\rho}$ is described by $\mathcal{D}_{BE_k}^{\check{}}$, and $\mathcal{D}_{BE_k}^{\hat{}}$ and $\mathcal{D}_{BE_k}^{\check{}}$ are corresponding up to depth $n-1$. By the inductive hypothesis, $\mathcal{D}_{BE_k}^{\hat{}}|_{k-1}$ and $\mathcal{D}_{BE_k}^{\check{}}|_{k-1}$ are corresponding up to depth $n-1$ (and, obviously, $\mathcal{D}_{BE_k}^{\check{}}|_{k-1}$ is a $\overline{B}$-successor of $\mathcal{D}'_{BE_k}|_{k-1}$).

Finally (only for cases with $k \geq 2$), let us consider a subtree of depth $k-2$ linked to the root of $\mathcal{D}_{BE_k}|_{k-1}$ via a $B$-edge. In this case, there exists (at least) a subtree of $\mathcal{D}_{BE_k}$, say $\mathcal{S}_{k-1}$, such that $\mathcal{S}_{k-1}|_{k-2}$ is the considered subtree of $\mathcal{D}_{BE_k}|_{k-1}$. Since $\mathcal{D}_{BE_k}$ and $\mathcal{D}'_{BE_k}$ are corresponding up to depth $n$, there exists a subtree $\mathcal{S}'_{k-1}$ of $\mathcal{D}'_{BE_k}$, connected to the root of $\mathcal{D}'_{BE_k}$ via a $B$-edge, corresponding to $\mathcal{S}_{k-1}$ up to depth $n-1$. By the inductive hypothesis $\mathcal{S}_{k-1}|_{k-2}$ and $\mathcal{S}'_{k-1}|_{k-2}$ are corresponding up to depth $n-1$ (the latter is a subtree of $\mathcal{D}'_{BE_k}|_{k-1}$ connected to the root of $\mathcal{D}'_{BE_k}|_{k-1}$ via a $B$-edge).

The remaining cases can be dealt with analogously. □